\documentclass{article}
\usepackage[preprint]{neurips_2026}

\usepackage{comment}
\usepackage[utf8]{inputenc} 
\usepackage[T1]{fontenc}    
\usepackage{hyperref}       
\usepackage{url}            
\usepackage{booktabs}       
 \usepackage{graphicx}
 \usepackage{amsmath}
 \usepackage{float}
 \usepackage{algorithm}
 \usepackage{amssymb}
\usepackage{algorithmic}
 \usepackage{amsfonts}       
\usepackage{nicefrac}       
\usepackage{microtype}      
\usepackage{xcolor}         

\newcommand{\xxb}[1]{\textcolor{blue}{#1}}
\usepackage[normalem]{ulem}
 
\title{Evolutionary foraging in grids: Intermittent search dynamics emerge in finite, depletable landscapes}

\author{%
Shailendra Bhandari\\
  Department of Computer Science\\
  OsloMet -- Oslo Metropolitan University\\
  St.~Olavs plass, N-0130 Oslo, Norway \\
  \texttt{shailendra.bhandari@oslomet.no} \\
\And
Alex Szorkovszky \\
Simula Research Laboratory\\ Numerical Analysis and Scientific Computing \\
Oslo, 0164, Norway \\
 \texttt{alex@simula.no} \\
\And
  Anis Yazidi \\
  Department of Computer Science\\
  OsloMet -- Oslo Metropolitan University\\
  St.~Olavs plass, N-0130 Oslo, Norway \\
  Department of Informatics\\ University of Oslo \\
  Gaustadall\'een 23B, 0373 Oslo, Norway \\
\texttt{anisy@uio.no} \\
\And
  Pedro G. Lind \\
  Department of Computer Science\\
  OsloMet -- Oslo Metropolitan University\\
  St.~Olavs plass, N-0130 Oslo, Norway \\
  Department of Technology\\ Kristiania University of Applied Sciences \\
  Oslo, Norway \\
  \texttt{pedrolin@oslomet.no} 
}

\begin{document}
\maketitle
\begin{abstract}
How search strategies evolve in finite, depletable landscapes remains a central question in foraging theory. We study this problem with an evolutionary simulation in which agents forage on a two-dimensional toroidal lattice containing non-renewable resources distributed either uniformly or as L\'evy dust. Each agent carries a heritable genome encoding step lengths, velocities, and turning angles, and selection acts on a fitness function combining energetic gain, movement cost, and coverage efficiency. By allowing movement traits to evolve without imposing a prescribed power-law step-length distribution, we test whether the evolved trajectories are better described by intermittent-search or L\'evy-walk dynamics. Strikingly, our results indicate that, in the finite depletion-driven landscapes considered here, evolved search is more consistent with intermittent dynamics than with strict scale-free L\'evy motion. To characterize the effective dynamics of the evolved trajectories, we take the average adjusted coefficient of determination when fitting second- and fourth-order displacement moments to intermittent search and L\'evy-walk models. While a L\'evy-like random walk fits with the numerical results from the evolutionary search very well (mean adjusted \(\bar{R}^2>0.9\) in most tested conditions), the intermittent search achieves a closer fit, with mean adjusted coefficients of determination(\(\bar{R}^2>0.99\)) for all tested resource distributions. This preference holds across the tested grid sizes and resource densities. Five independent evolutionary runs per environment on a $503\times503$ grid at nominal resource density $\rho=0.15$ reproduce this preference for the uniform environment and the five L\'evy-dust environments. Evolution rapidly reshapes the movement genome toward short displacements while retaining a sparse tail of longer relocations, consistent with local exploitation punctuated by occasional transfer. The framework provides a controlled setting for studying how search rules emerge under resource limitation and may inform resource-constrained exploration in autonomous systems. Project webpage: \url{https://evo-foraging.github.io/}
\end{abstract}

\section{Introduction}
\label{introduction}
Foraging efficiency is a central problem in foraging theory, because survival and reproductive success often depend on how organisms balance resource gain against movement cost in uncertain environments \cite{CHARNOV1976129,mobbs2018foraging,Wang2025ForagingEvolution,Chaumont2016SustainedForaging}. A major line of work has focused on the L\'evy walk (LW) foraging hypothesis, which proposes that power-law-distributed step lengths can optimize search in sparse landscapes \cite{viswanathan1999optimizing,viswanathan2011physics,RevModPhys.87.483}. The inverse-square case  \(p(s)\propto s^{-2}\), often referred to as the Cauchy walk, is a particularly important benchmark because it corresponds to the classical exponent near \(2\) associated with efficient L\'evy foraging \cite{doi:10.1126/sciadv.abe8211}. Empirical and theoretical studies have reported L\'evy-like movement across many biological systems, including marine predators, birds, insects, and human search behavior \cite{sims2020marine,LuisaVissat2023,gautestad2013levy,10.1177/26339137241228858}. L\'evy-like movement has also influenced algorithmic search and optimization, where rare long relocations can improve exploration in high-dimensional or weakly informative objective landscapes \cite{doi:10.5772/60414,leon2021mutations,HE2023376}.

The generality of the LW foraging hypothesis remains debated. Its optimality depends on assumptions about spatial scale, resource renewal, target density, memory, and boundary effects \cite{10.1098/rsif.2014.1158,REYNOLDS2006384,PhysRevResearch.6.023274}. In finite landscapes, very long moves can be altered by boundaries or by repeated crossings of already explored regions. In depletable landscapes, local resource consumption alters the effective environment during the search, forcing the forager to balance exploitation of nearby resources against relocation to less-depleted areas. Moreover, L\'evy-like trajectories may arise either from intrinsic movement rules or from interactions between the forager and a structured environment, making it difficult to infer the underlying strategy from step-length statistics alone \cite{gautestad2013levy,doi:10.1086/729220}. 
A closely related alternative is an intermittent search (IS), in which movement alternates between localized search phases and relocation phases \cite{RevModPhys.83.81}. Unlike a strict LW, which is defined by a scale-free move-length law, IS can generate bimodal or multimodal movement statistics, with frequent local displacements and less frequent long transfers. Recent evolutionary simulations that relaxed the assumption of power-law move lengths found that IS can evolve de novo and outcompete LWs under a broad class of foraging conditions \cite{doi:10.1086/729220,wosniack2022maintenance}. These results suggest that selection may favor a structured mixture of local exploitation and relocation rather than a strict scale-free movement law. Evolutionary models have also supported the idea that L\'evy-like search can arise under sparse-resource conditions \cite{wosniack2022maintenance,10.1371/journal.pcbi.1005774}. Previous work has connected stochastic-search models with classical foraging theory and examined the assumptions and competing interpretations of the L\'evy-foraging hypothesis \cite{Bartumeus_2009,Klages2023}. Within this broader context, the present study considers a specific evolutionary setting with finite, depletable landscapes and compares the resulting trajectories using LW and IS moment models.

Recent random-walk theory also provides tools for studying foraging under depletion. Range-controlled random walks, in which movement rates depend on the number of distinct sites already visited, show how exploration dynamics can change when motion is coupled to accumulated coverage \cite{redner2001guide,10.1063/1.1704269,PhysRevLett.130.227101}. Starving random-walk models make this connection explicit by asking whether a forager can discover a new food-containing site before a metabolic deadline \cite{PhysRevLett.113.238101,ben-Avraham_Havlin_2000,Havlin01011987,PhysRevLett.132.127101}. In this setting, the new-site discovery interval \(\tau_n\), defined as the waiting time between discovery of the \(n\)-th and \(n+1\)-th distinct lattice sites, provides a natural measure of exploration under depletion. These ideas motivate the analysis of waiting-time structure in addition to spatial trajectory shape.

Distinguishing LW and IS dynamics requires more than inspecting apparent heavy tails. Moment-based approaches provide one route to this distinction. In recent work, the second and fourth displacement moments were shown to separate LW and IS dynamics because LWs produce scale-dependent moment growth governed by the L\'evy exponent, whereas IS produces a two-timescale structure associated with relocation and local search phases \cite{2bzm-t9k1,BHANDARI2025102334}. This provides a quantitative way to test whether evolved trajectories are better described by strict LW dynamics or by IS.

Here, we develop an evolutionary simulation framework that combines genome-encoded movement traits, finite lifetimes, explicit resource depletion, and controlled resource clustering. Agents forage on two-dimensional toroidal lattices containing non-renewable resources distributed either as L\'evy dust or uniformly. Selection acts on a fitness function that rewards energetic efficiency and low-redundancy spatial coverage. We analyze how movement genomes evolve across resource environments, classify evolved trajectories using second- and fourth-moment fits to LW and IS models, and quantify waiting-time statistics under depletion. This framework allows us to ask whether adaptive search in finite, depletable landscapes converges toward strict scale-free LW dynamics or toward IS, combining local exploitation with occasional relocation.
\section{Methods}
\subsection{Resource landscapes and depletion}
\label{sec:model_components}
We simulate evolutionary foraging on a discrete two-dimensional torus of size \(L\times L\), populated by depletable unit-valued resources. Clustered resource landscapes were generated as L\'evy-dust fields, following standard constructions in which successive resource placements are separated by power-law-distributed distances \cite{viswanathan2011physics,doi:10.1086/729220,burrough1981fractal}. Resources are non-renewable: once encountered by an agent, the resource at that site is removed for the remainder of that lifetime. The realized trajectory is therefore shaped jointly by the initial spatial resource field and by depletion during search. Resource placement follows a random walk on the torus whose placement-step lengths \(\ell\) are sampled from
\begin{equation}
p_{\mu}(\ell)
=
\frac{\ell^{-\mu}}
{\sum_{k=1}^{\lfloor L/2\rfloor} k^{-\mu}},
\qquad
\ell=1,2,\dots,\lfloor L/2\rfloor .
\label{eq:levy_pmf}
\end{equation}
Directions are sampled uniformly, and positions are wrapped modulo \(L\), corresponding to periodic boundary conditions. Each reached lattice site is marked as containing a unit resource. The placement walk is run for \(A_\rho=\lfloor \rho L^2\rfloor\) placement attempts. Since the placement walk may return to sites that have already been occupied, repeated hits do not create additional occupied sites. Thus, for L\'evy-dust landscapes, \(\rho\) specifies the nominal number of placement attempts per lattice site, whereas the realized density is the number of distinct occupied sites divided by \(L^2\). The realized density can therefore differ from the nominal value, especially when repeated placements occur.
\begin{figure}[t]
\centering
\includegraphics[width=\textwidth]{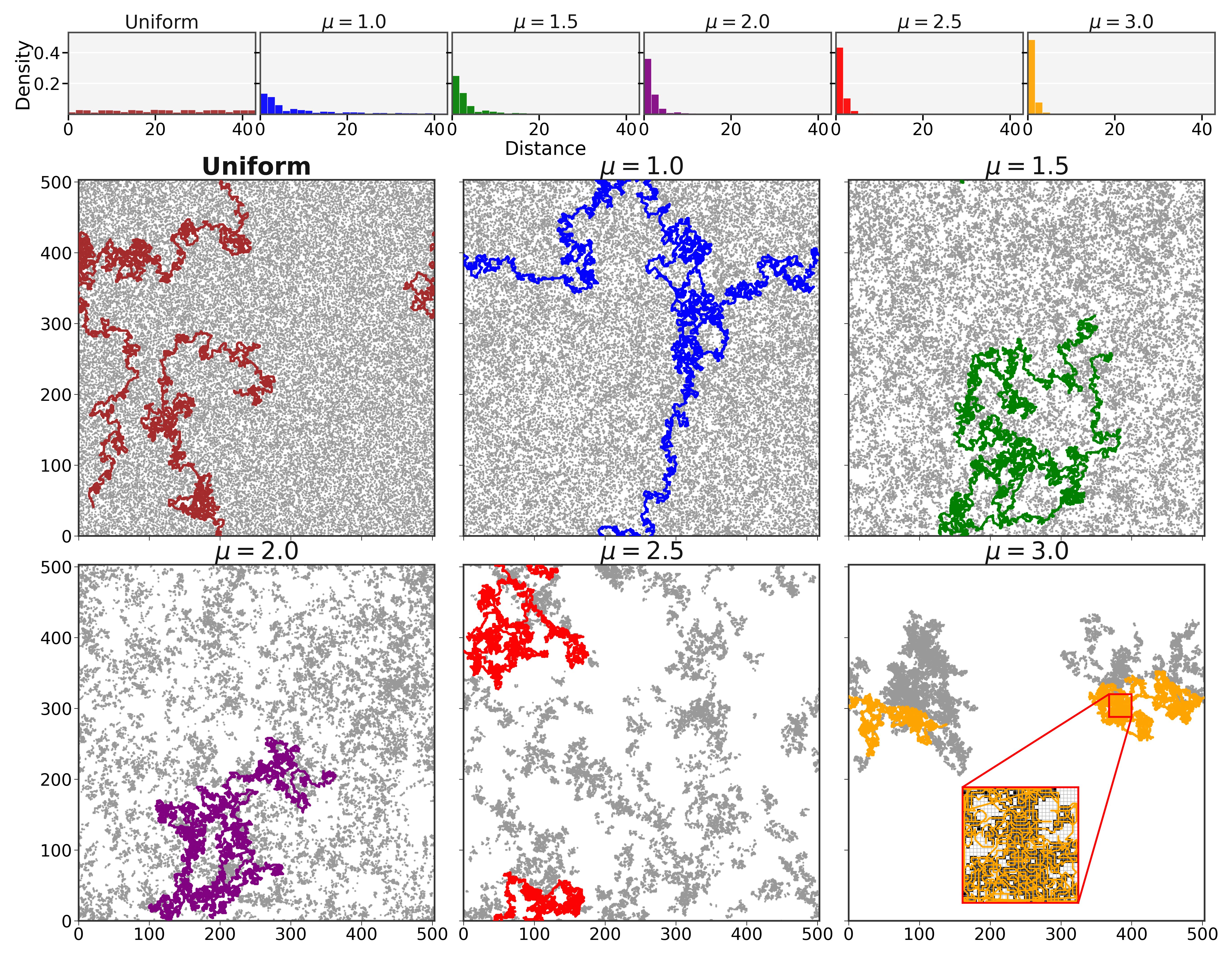}
\caption{\protect
Resource landscapes and best evolved trajectories for the \(L=503\) torus.
\textbf{Top row}: placement-step distributions used to generate the resource fields, shown for the uniform control and for L\'evy-dust environments with \(\mu\in\{1.0,1.5,2.0,2.5,3.0\}\).
\textbf{Main panels}: resource locations (gray points) overlaid with the trajectory of the best evolved agent in each environment (colored line).
Changing \(\mu\) changes the spatial correlation structure of the resource field by altering the frequency of long placement steps: smaller \(\mu\) gives heavier-tailed placement steps, whereas larger \(\mu\) suppresses long placement steps.
The inset in the \(\mu=3.0\) panel shows a magnified local patch.
}
\label{fig:levy_examples}
\end{figure}
We consider L\'evy-dust exponents $\mu\in\{1.0,1.5,2.0,2.5,3.0\}$, which control the spatial correlation of resource placement. Smaller \(\mu\) values allow more long placement steps and produce more spatially dispersed resource fields, whereas larger \(\mu\) values suppress long placement steps and increase local aggregation. As a reference condition, we also use a uniform environment generated by an independent Bernoulli occupation of lattice sites with probability \(\rho\). In the L\'evy-dust fields, repeated placements at the same site leave the occupancy state unchanged. The simulations used in this study consider three linear system sizes, $L\in\{211,503,1009\}$. At fixed nominal density $\rho$, changing $\mu$ alters both spatial clustering and realized resource abundance through repeated resource placements. Comparisons across $\mu$ therefore reflect changes in both resource geometry and abundance. Table~\ref{tab:realized_occupancy} in Appendix \ref{app:robustness_results} reports the realized occupancies of the saved pre-depletion maps for $L=503$ at nominal densities $\rho\in\{0.15,0.30,0.45\}$. For each grid size, nominal resource density, and resource environment, the resource field is generated on the torus and used as the initial landscape for the corresponding evolutionary run. Figure~\ref{fig:levy_examples} shows representative resource maps and best evolved trajectories for the \(L=503\) system, together with the placement-step distributions used to generate the L\'evy-dust environments. Additional implementation details with the simulation parameters are provided in Appendix~\ref{app:evolutionary_algorithm}, Table~\ref{tab:simulation_parameters}.
\subsection{Agent movement and genetic representation} \label{sec:agent}
Each agent carries three heritable integer-valued genomes of fixed length \(G\): a step-length genome, a velocity genome, and a turn-angle genome. The admissible values are
$s\in\{1,\;\ldots,\;\lfloor L/2\rfloor\}, \;
v\in\{1,\;\ldots,\;L\},\;
\kappa\in\{0,\;\ldots,\;n_\theta-1\}$, with \(n_\theta=36\), where \(\kappa\) maps to the turning angle \(\theta_\kappa = 10^\circ \kappa\). Each genome contains \(G=30\) entries. The genomes are represented as finite lists of admissible trait values rather than as explicit parametric distributions. At each movement decision, the agent samples one entry uniformly from each list. Selection, recombination, and mutation, therefore, act on the empirical frequencies of encoded trait values, allowing the movement statistics to emerge from the evolving genome. 
At the beginning of a lifetime, the agent is initialized at a resource-bearing lattice site chosen uniformly at random. If no resource-bearing site is available, the initial position is drawn uniformly from the lattice. The initial heading is sampled uniformly. Motion then proceeds on the periodic lattice for the prescribed lifespan \(T_{\mathrm{life}}\). At each movement decision, the intended step length and velocity are sampled from the corresponding genomes. With probability \(p_{\mathrm{cue}}=0.5\), the cue branch searches a square window of half-width \(r_{\mathrm{cue}}=60\), clipped at the lattice boundaries. Movement is permitted across the periodic boundary, but sensing does not wrap. If the window contains an undepleted resource, the heading is set directly toward the nearest one by Euclidean distance. Otherwise, a uniformly random heading is selected without falling through to the genome rule. With probability \(1-p_{\mathrm{cue}}\), the inherited turning increment is applied. Thus, \(p_{\mathrm{cue}}\) is the branch-selection probability, while the frequency of resource-directed turns also depends on resource availability. This rule introduces limited local environmental information without giving the agent global knowledge of the resource field \cite{NOLTING2015126}.

The intended displacement is a straight movement segment of length \(s\), realized on the lattice along the current heading and subject to periodic wrapping. The velocity determines the maximum distance advanced during one movement update within that segment. If \(s_{\mathrm{rem}}\) denotes the remaining segment length, movement update \(j\) advances \(\Delta_j=\min(v,s_{\mathrm{rem}})\), followed by \(s_{\mathrm{rem}}\leftarrow s_{\mathrm{rem}}-\Delta_j\). An uninterrupted segment therefore requires \(\lceil s/v\rceil\) movement updates. The lifespan \(T_{\mathrm{life}}=8000\) counts movement updates, and the accumulated movement cost is \(Q(t)=c_{\mathrm{move}}\sum_{j=1}^{t}\Delta_j\), where \(c_{\mathrm{move}}=0.5\). Thus, \(\Delta_j\) is the scalar advance used to reduce the remaining segment length and to calculate movement cost. It need not equal the Euclidean length of the rounded, pre-wrap coordinate increment. A site counts as visited only when it is the lattice position at the end of a movement update. Sites between consecutive endpoints do not increment \(N(t)\) or \(R(t)\), and resources on those sites are neither encountered nor consumed. If a resource is present at an update endpoint, the resource is consumed, the cumulative harvested energy \(E_{\mathrm{res}}(t)\) is incremented by the prescribed gain, and the current movement segment is terminated at that endpoint. During each lifetime, we record the cumulative movement cost \(Q(t)\), the number \(N(t)\) of distinct lattice sites visited, and the revisit count \(R(t)\). The initial position is included as the first element of the endpoint sequence, and all coverage and waiting-time statistics are calculated from this same sequence. These quantities serve as the state variables in the fitness function defined below.
\subsection{Fitness evaluation} \label{sec:fitness}
Fitness is defined to reward agents that acquire resources at low movement cost while maintaining broad, low-redundancy spatial coverage. We choose the product of an energetic-efficiency term and a coverage-efficiency term as the modelling objective, namely
\begin{equation}
\mathcal F(t)
=
\eta_E(t)\,\eta_C(t) .
\label{eq:fitness}
\end{equation}
To define the energetic efficiency \(\eta_E(t)\), we consider the cumulative movement cost \(Q(t)\), calculated from the scalar movement increments $\Delta_j$, and the harvested resource energy \(E_{\mathrm{res}}(t)\). This energy is computed as $E_{\mathrm{res}}(t)=\varepsilon n_{\mathrm{res}}(t)$, where \(\varepsilon\) is the gain from one resource encounter and \(n_{\mathrm{res}}(t)\) is the number of resources collected up to time \(t\). The energetic efficiency is $\eta_E(t) ={E_{\mathrm{res}}(t)}/ ({E_{\mathrm{res}}(t)+Q(t)}) = 1/({1+Q(t)/E_{\mathrm{res}}(t)})$. When \(E_{\mathrm{res}}(t)=0\), we set \(\eta_E{(t)}=0\), so that agents that collect no resources receive zero energetic efficiency. Note that $0\le \eta_E(t)\le 1$. This term is large when the agent obtains substantial reward at low locomotion cost and decreases when search becomes energetically inefficient.

To quantify coverage-efficiency, let \(N(t)\) be the number of distinct lattice sites visited up to time \(t\), and let \(R(t)\) be the number of revisits to sites that have already been visited. Then, similarly to $\eta_E(t)$ we define
$\eta_C(t)
= N(t) /( N(t)+R(t) ) = 1 /( 1 + R(t)/N(t) )$ with
$0\le \eta_C(t)\le 1$.
This term approaches one when most visits are to previously unvisited sites and decreases when the trajectory becomes dominated by returns to already explored locations. 
Combining the two terms gives
\begin{equation}
\mathcal F(t)
=
\frac{1}{1+\tfrac{Q(t)}{E_{\mathrm{res}}(t)}}
\cdot
\frac{1}{1+\tfrac{R(t)}{N(t)}} .
\label{eq:fitness_expanded}
\end{equation}
By construction, \(0\le \mathcal F(t)\le 1\). High fitness, therefore, requires both profitable energy acquisition and efficient spatial coverage. In the simulations, the quantities entering Eq.~\eqref{eq:fitness_expanded} are tracked directly during each trajectory. The harvested energy \(E_{\mathrm{res}}(t)\) is updated whenever a resource is encountered and consumed. The cumulative cost \(Q(t)\) is incremented according to the realized motion under the fixed movement-cost rule. Distinct visited sites are recorded explicitly, and each return to an already visited site increments the revisit count \(R(t)\). Fitness is evaluated at the end of the lifetime and used for fitness-proportionate reproduction with elitism, recombination, and bounded Gaussian mutation.
\subsection{Classifying evolved trajectories: intermittent or L\'evy?} \label{classification}
To determine whether the evolved trajectories are better described by an IS model or by an LW model, we fitted theoretical displacement moments to empirical moments computed from the simulated paths. The procedure follows the log-moment classification framework introduced in Ref.~\cite{2bzm-t9k1}, with fitting and scoring routines implemented in the IntLevPy analysis library \cite{BHANDARI2025102334}. For each trajectory, we computed displacements over lag \(t_s\), $\Delta \mathbf r(t;t_s) = \mathbf r(t+t_s)-\mathbf r(t)$, using unwrapped coordinates to avoid artificial shortening at periodic boundary crossings. The empirical radial moments were
\begin{equation}
m_k^{\mathrm{obs}}(t_s)
=
\left\langle
\|\Delta \mathbf r(t;t_s)\|^k
\right\rangle_t,
\qquad
k\in\{2,4\},
\label{eq:emp_mom_revised}
\end{equation}
where the average is taken over all admissible time origins \(t\) along the trajectory. Fitting used a set of \(n_{\mathrm{lag}}=28\) unique lag times \(t_s\), generated as \(t_s=\lfloor2^{n/10}\rfloor\) for \(n=0,\ldots,53\), with all lag times measured in movement updates.

The LW model is defined by straight flights at constant speed, with flight durations drawn from a heavy-tailed distribution,
\begin{equation}
\psi(\tau_{\mathrm f})
=
\frac{\gamma}{\tau_0}
\left(
1+\frac{\tau_{\mathrm f}}{\tau_0}
\right)^{-(1+\gamma)} ,
\label{eq:lw_pdf_revised}
\end{equation}

parameterized by \((v_L,\gamma,\tau_0)\). The IS model alternates between a diffusive phase of diffusivity \(D\) and a ballistic phase of speed \(v_B\), with Poisson switching rates \(\lambda_{BD}\) and \(\lambda_{DB}\). For the IS model, explicit analytical expressions for the second and fourth displacement moments are available \cite{2bzm-t9k1}. See Appendix \ref{append:moments}. For the LW model, the corresponding moment behavior in log-space was evaluated within the same fitting framework used in Ref.~\cite{2bzm-t9k1} and in the associated implementation software library \cite{BHANDARI2025102334}. For each model, parameters were estimated by minimizing the mean squared discrepancy in logarithmic moment space,
\begin{equation}
\mathcal J
=
\frac{1}{2n_{\mathrm{lag}}}
\sum_{t_s}
\sum_{k\in\{2,4\}}
\left[
\log m_k^{\mathrm{obs}}(t_s)
-
\log m_k^{\mathrm{mod}}(t_s)
\right]^2 ,
\label{eq:loss_revised}
\end{equation}
subject to the corresponding physical parameter constraints. Working in log-moment space reduces the dominance of large-amplitude moment values and gives a more balanced comparison across lag scales. Goodness of fit was quantified using the adjusted coefficient of determination computed in logarithmic moment space, accounting for the three free parameters of LW and four of IS. 
For each trajectory and model, the adjusted \(R^2\) values for \(m_2\) and \(m_4\) were averaged to give \(\bar R^2_{\mathrm{IS}}\) and \(\bar R^2_{\mathrm{LW}}\). We then defined the model-comparison score
\begin{equation}
\Gamma=\bar R^2_{\mathrm{IS}}-\bar R^2_{\mathrm{LW}} .
\label{eq:gamma_revised}
\end{equation}
Thus, $\Gamma>0$ indicates that the IS model provides a better moment-level description of the trajectory, whereas $\Gamma<0$ favors the LW model. The classification is therefore a comparison between two idealized dynamical descriptions, not evidence that agents explicitly implement either model; summary statistics are computed over stored generation-wise best-agent trajectories. The classification concerns realized trajectories shaped jointly by the genomes, sensory cue, depletion, and encounter-triggered stopping. Synthetic-trajectory classification and IS parameter recovery are reported in Appendix~\ref{app:synthetic_validation}.

\section{Results: evolved search dynamics}\label{results}
We analyzed evolved trajectories and movement traits after 1500 generations of selection across all resource environments. Across conditions, fitness increased early in evolution, step-length genomes shifted toward short displacements, and the evolved trajectories were better approximated by the IS model than by the strict LW model. We first describe the spatial trajectories and evolutionary changes in the genomes, and then compare the evolved trajectories with the IS and LW models using the second and fourth-order displacement moments.
\begin{figure}[t]
\centering
\fbox{\includegraphics[width=0.44\textwidth]{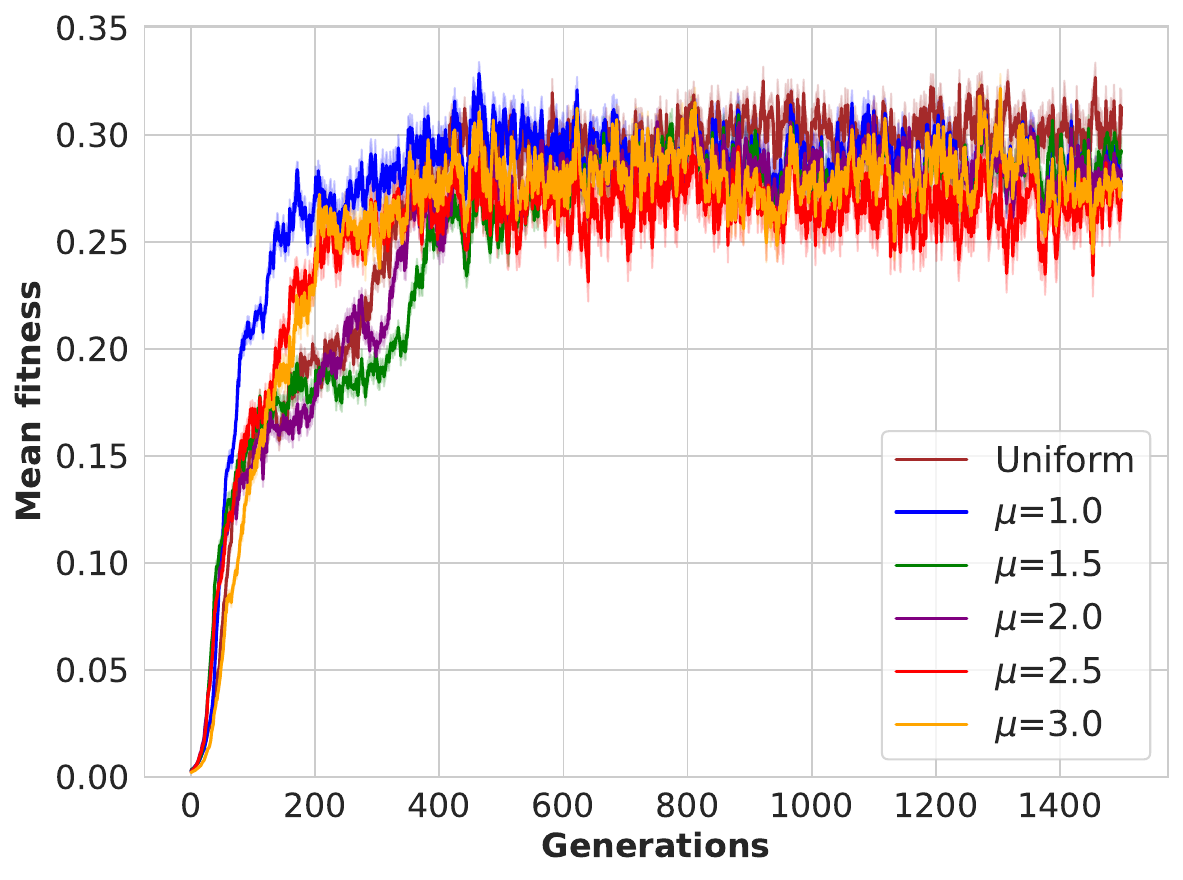}\hfill
\includegraphics[width=0.53\textwidth]{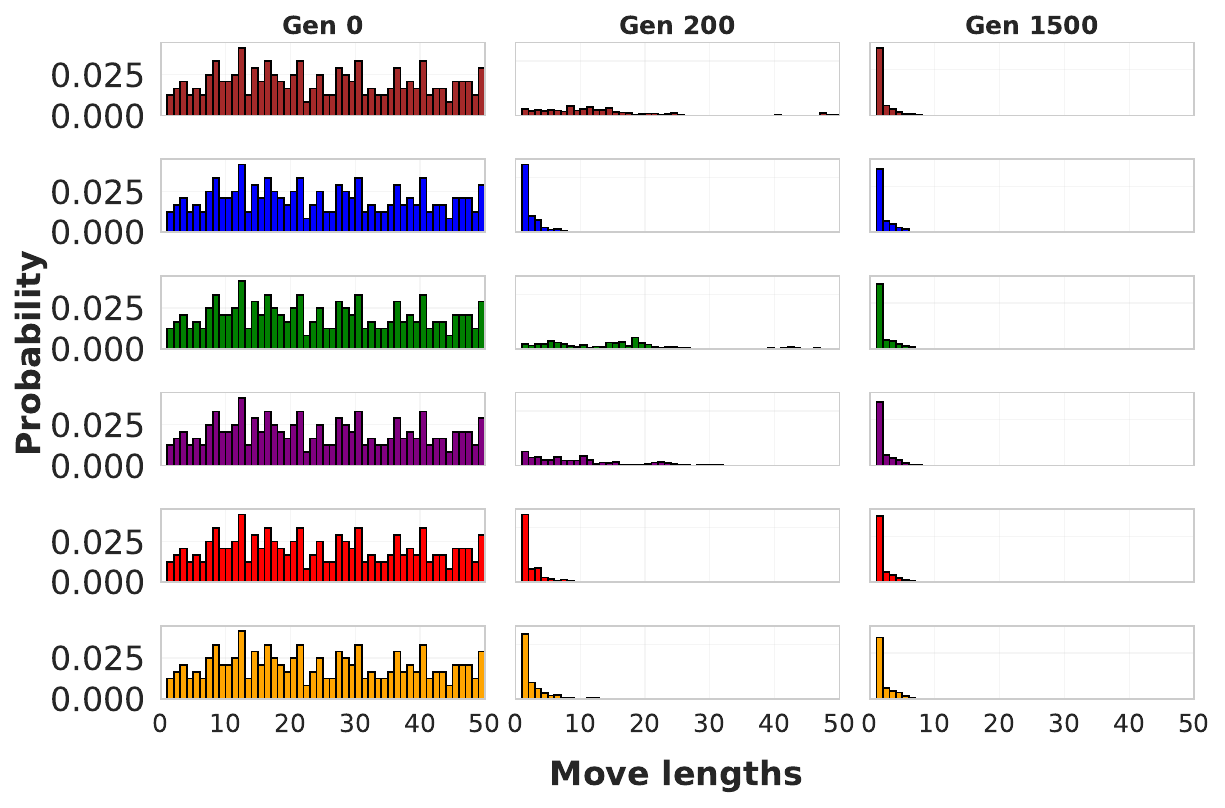}}
\caption{
Evolution of fitness and the step lengths on the \(503\times503\) grid.
\textbf{Left:} mean fitness over 1500 generations for each resource environment, with shaded standard errors across agents within a single evolutionary run at each generation. \textbf{Right:} distributions of step-length genome entries at generations 0, 200, and 1500.
Selection rapidly concentrates the encoded step-length pool at short values while retaining a sparse tail of longer entries.}
\label{fig:evolution}
\end{figure}
The trajectories in Figure~\ref{fig:levy_examples} show the best evolved individual on the \(503\times503\) torus, overlaid on the corresponding resource maps with $\rho=0.15$. The overlays show spatially heterogeneous movement, with dense local path segments and occasional larger displacements visible in several environments. We treat this as a qualitative description of the trajectories; the quantitative distinction between LW and IS dynamics is made using second- and fourth-order displacement moments. As the resource field changes with \(\mu\), the spatial organization of the trajectories changes as well, reflecting the interaction between the evolved movement genome, resource depletion, and movement termination upon resource encounter. Because resource encounters can truncate planned movement segments, realized displacement statistics need not coincide with the raw genome-encoded step-length lists.

Figure~\ref{fig:evolution} summarizes the evolutionary dynamics on the \(503\times503\) grid. Mean fitness rises rapidly during the early generations and then fluctuates around an environment-dependent plateau. The curves do not show a simple monotonic ordering with \(\mu\), indicating that final performance reflects the combined effects of resource geometry, depletion, local cueing, and movement cost. The genome histograms show a clearer and more robust trend: initially broad step-length lists become concentrated at short steps by generation 200 and remain strongly short-biased at generation 1500, while retaining a sparse tail of longer entries. These histograms describe the encoded sampling pool from which step lengths are drawn, not the realized displacement distribution along the path. A complementary stabilization analysis is provided in Appendix~\ref{sec:fitness_appendix}. There, the best-fitness trajectories are shown together with the Jensen--Shannon divergence \(\mathrm{JS}(P_g(s),P_0(s))\) of the population-level step-length distribution from its initial state. Both quantities show rapid early change followed by an operational plateau, with plateau generations estimated between \(g=390\) and \(g=547\) on the \(503\times503\) grid using a rolling-slope criterion.

\begin{figure}[t]
\centering
\includegraphics[width=\textwidth]{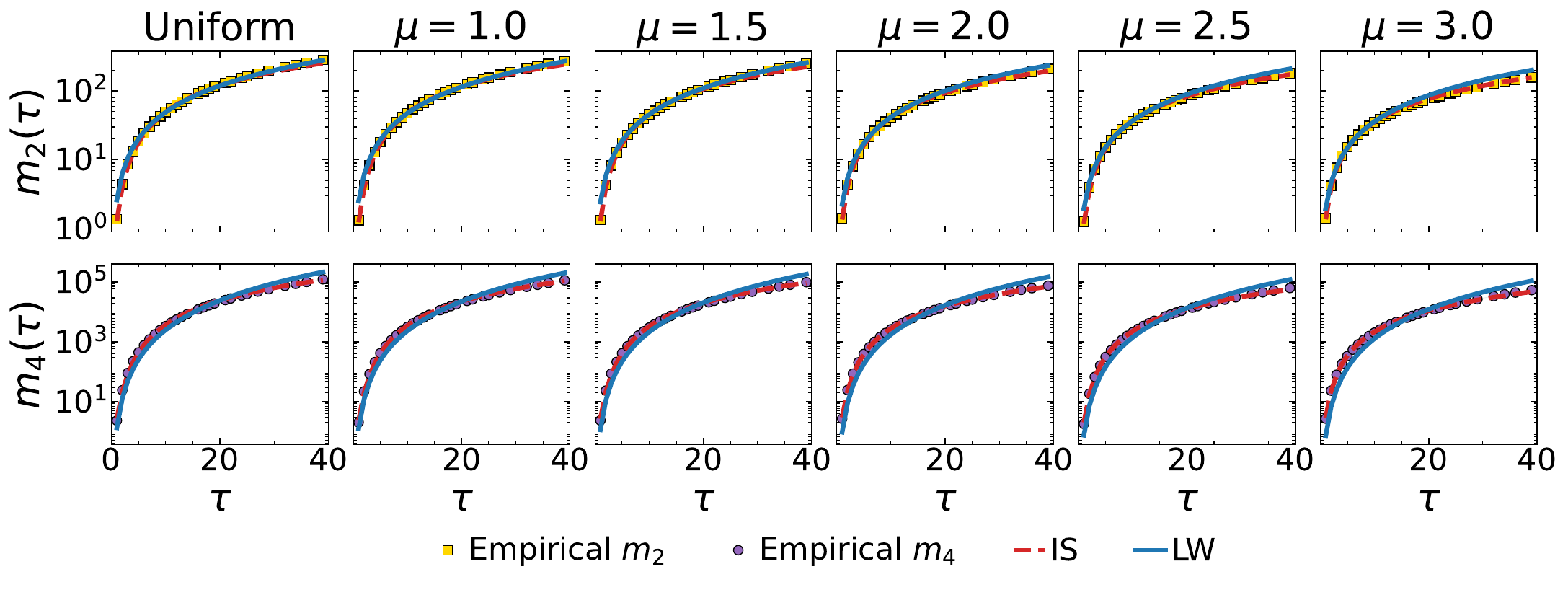}
\caption{Moment-based comparison of evolved trajectories on the \(503\times503\) grid.
Empirical second moments \(m_2\) and fourth moments \(m_4\) are compared with best-fit IS and LW models. Columns correspond to the six resource environments. Across environments, IS gives a closer fit to the empirical moment curves.}
\label{fig:evo_moments}
\end{figure}

Figure~\ref{fig:evo_moments} compares empirical second and fourth displacement moments with the best-fit IS and LW models for the \(503\times503\) grid ($\rho=0.15$). The two models often appear close at the level of the second moment, but differences are clearer in the fourth moment, which is more sensitive to rare large displacements. Across all environments, the IS model provides a closer moment-level description. This visual pattern is consistent with the quantitative model-comparison scores in Table~\ref{tab:model_comparison_summary} for every environment and grid size, \(\Gamma=\bar R^2_{\mathrm{IS}}-\bar R^2_{\mathrm{LW}}>0\). Thus, within this moment-based classification framework, the evolved trajectories are better approximated by the IS than by the strict LW model. At \(L=503\) and \(\rho=0.15\), five independent evolutionary runs per environment reproduced the preference for IS, with \(\Gamma>0\) in all 30 run--environment combinations. Positive \(\Gamma\) was also observed across all six environments in the single-run density sweep at \(\rho\in\{0.005,0.010,0.15,0.30,0.45\}\) (Appendix~\ref{app:density_results}, Fig.~\ref{fig:density_results_503}). 

Re-evolution under an ablated objective \(F=\eta_E\) retained \(94.7\%\)--\(100.1\%\) of the baseline coverage efficiency and yielded \(\Gamma>0\) in all six environments (Appendix Table~\ref{tab:fitness_ablation}). Setting \(p_{\mathrm{cue}}=0\) during fixed-genome evaluation also preserved \(\Gamma>0\), although the margin decreased in the uniform and \(\mu\in\{1.0,1.5,2.0\}\) environments and increased for \(\mu\in\{2.5,3.0\}\) (Appendix Table~\ref{tab:cue_off_evaluation}). Because the genomes were not re-evolved without the cue, the latter analysis measures its effect on expressed trajectories rather than on evolutionary outcomes.

\begin{figure}[t]
    \centering
    \includegraphics[width=\linewidth]{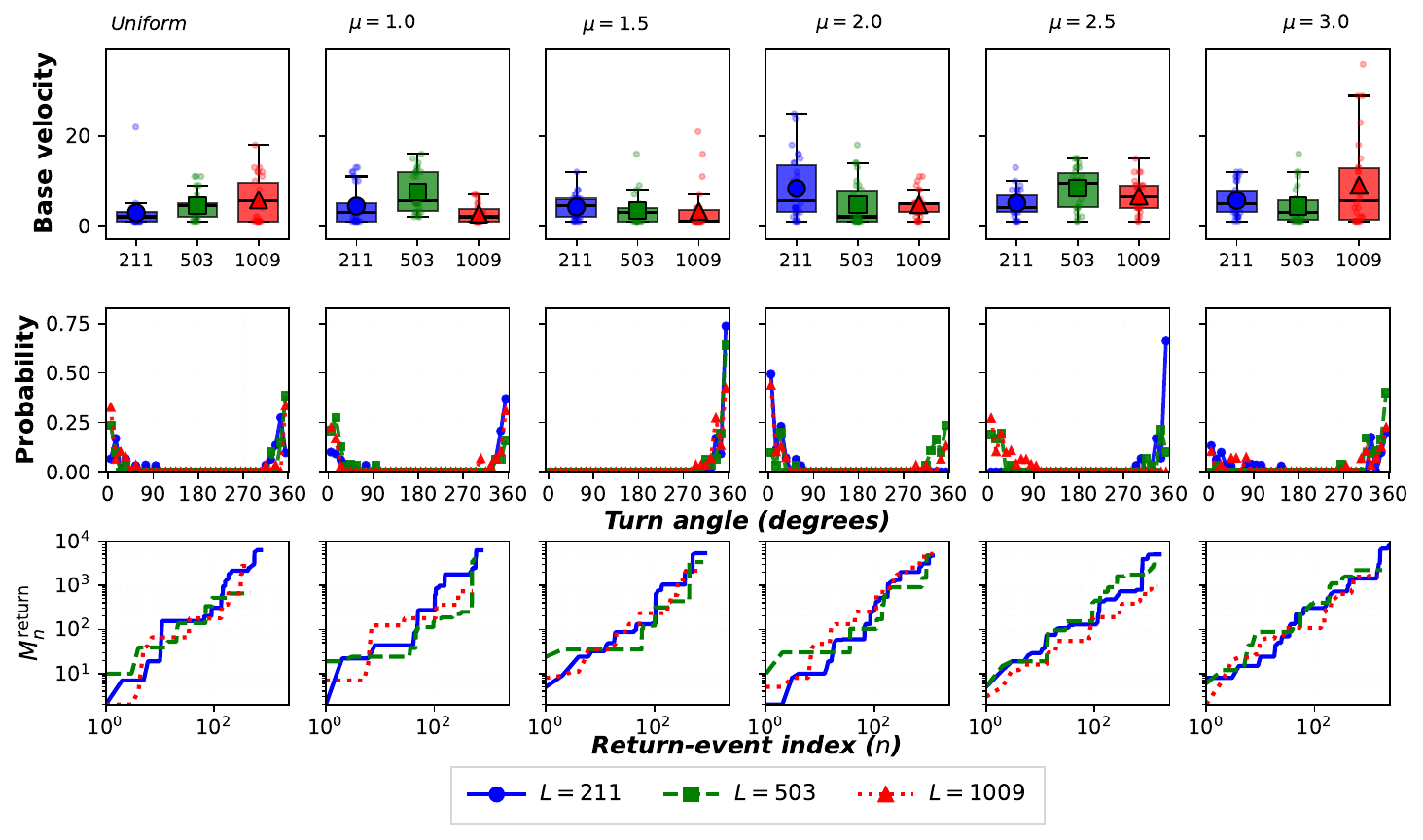}
    \caption{Movement traits and return-interval statistics of the best evolved agents across grid sizes after 1500 generations. Columns correspond to resource environments. \textbf{Top row:} velocity-genome entries of the best evolved agents for \(L=211\), \(503\), and \(1009\). \textbf{Middle row:} turn-angle increments recorded along the best evolved trajectories. \textbf{Bottom row:} running maximum of return intervals \(M_j^{\mathrm{return}}\) versus return-event index \(j\).}
    \label{fig:cross_size_traits}
\end{figure}

\begin{table}[t]
\centering
\caption{
Model-comparison summary across environments and grid sizes at \(\rho=0.15\). For each case, we report \(\Gamma=\bar{R}^2_{\mathrm{IS}}-\bar{R}^2_{\mathrm{LW}}\), together with the adjusted coefficients of determination. For \(L=211\) and \(L=1009\), values are mean \(\pm\) standard deviation over the final 500 best-agent trajectories from one evolutionary run. For \(L=503\), values are mean \(\pm\) standard deviation across five independent run-level means, each computed from the final 500 best-agent trajectories. The runs used independent resource maps and GA initializations. Positive \(\Gamma\) indicates preference for the IS model. All results are reported with significant figures in accordance with their respective standard deviations.
}
\label{tab:model_comparison_summary}
\begin{tabular}{llccc}
\toprule
Grid & Environment & \(\Gamma\) & \(\bar{R}^2_{\mathrm{IS}}\) & \(\bar{R}^2_{\mathrm{LW}}\) \\
\midrule
\(211\times211\) & uniform & \(0.028\pm0.004\) & \(0.9985\pm0.0004\) & \(0.970\pm0.004\) \\
\(211\times211\) & \(\mu=1.0\) & \(0.019\pm0.002\) & \(0.9984\pm0.0002\) & \(0.980\pm0.002\) \\
\(211\times211\) & \(\mu=1.5\) & \(0.022\pm0.003\) & \(0.9981\pm0.0003\) & \(0.976\pm0.003\) \\
\(211\times211\) & \(\mu=2.0\) & \(0.024\pm0.005\) & \(0.9993\pm0.0002\) & \(0.975\pm0.005\) \\
\(211\times211\) & \(\mu=2.5\) & \(0.030\pm0.007\) & \(0.9987\pm0.0005\) & \(0.969\pm0.007\) \\
\(211\times211\) & \(\mu=3.0\) & \(\mathbf{0.08\pm0.02}\) & \(0.9986\pm0.0006\) & \(0.92\pm0.02\) \\

\midrule

\xxb{\(503\times503\)} & \xxb{uniform} & \xxb{\(0.017\pm0.002\)} & \xxb{\(0.9983\pm0.0004\)} & \xxb{\(0.982\pm0.002\)} \\
\xxb{\(503\times503\)} & \xxb{\(\mu=1.0\)} & \xxb{\(0.015\pm0.005\)} & \xxb{\(0.9986\pm0.0004\)} & \xxb{\(0.984\pm0.005\)} \\
\xxb{\(503\times503\)} & \xxb{\(\mu=1.5\)} & \xxb{\(0.018\pm0.006\)} & \xxb{\(0.9987\pm0.0004\)} & \xxb{\(0.981\pm0.006\)} \\
\xxb{\(503\times503\)} & \xxb{\(\mu=2.0\)} & \xxb{\(0.030\pm0.002\)} & \xxb{\(0.9991\pm0.0002\)} & \xxb{\(0.969\pm0.002\)} \\
\xxb{\(503\times503\)} & \xxb{\(\mu=2.5\)} & \xxb{\(0.039\pm0.010\)} & \xxb{\(0.9992\pm0.0002\)} & \xxb{\(0.960\pm0.009\)} \\
\xxb{\(503\times503\)} & \xxb{\(\mu=3.0\)} & \xxb{\(\mathbf{0.077\pm0.032}\)} & \xxb{\(0.9991\pm0.0002\)} & \xxb{\(0.922\pm0.031\)} \\
\midrule
\(1009\times1009\) & uniform & \(0.017\pm0.001\) & \(0.9983\pm0.0003\) & \(0.981\pm0.001\) \\
\(1009\times1009\) & \(\mu=1.0\) & \(0.019\pm0.001\) & \(0.9985\pm0.0002\) & \(0.979\pm0.001\) \\
\(1009\times1009\) & \(\mu=1.5\) & \(0.022\pm0.006\) & \(0.9987\pm0.0006\) & \(0.977\pm0.006\) \\
\(1009\times1009\) & \(\mu=2.0\) & \(0.031\pm0.005\) & \(0.9992\pm0.0003\) & \(0.968\pm0.004\) \\
\(1009\times1009\) & \(\mu=2.5\) & \(0.06\pm0.02\) & \(0.9993\pm0.0002\) & \(0.94\pm0.02\) \\
\(1009\times1009\) & \(\mu=3.0\) & \(\mathbf{0.11\pm0.04}\) & \(0.9990\pm0.0006\) & \(0.89\pm0.04\) \\

\bottomrule
\end{tabular}
\end{table}
Figure~\ref{fig:cross_size_traits} compares movement traits and return-interval statistics of the best evolved agents across the three grid sizes. The top row shows velocity-genome entries. The median velocities range from \(1.0\) to \(9.5\) lattice units per update across the displayed conditions, indicating that the characteristic speed scale is not fixed universally, but varies with resource environment and system size. The middle row shows the distribution of turn-angle increments recorded along the best evolved trajectories. Angles are represented modulo \(360^\circ\) and binned in \(10^\circ\) intervals. Probability mass near \(0^\circ\) or \(360^\circ\) corresponds to nearly forward motion, whereas mass at intermediate angles corresponds to reorientation events. The bottom row shows the running maximum of return intervals, $M_j^{\mathrm{return}}=\max\{\Delta t^{\mathrm{return}}_1,\;\ldots,\;\Delta t^{\mathrm{return}}_j\}$, as a function of the return-interval index \(j\). Here \(\Delta t_j^{\mathrm{return}}\) denotes the \(j\)-th observed return interval between successive visits to the same lattice site. The staircase form is expected for a record process: \(M_j^{\mathrm{return}}\) changes only when a newly observed return interval exceeds all previous return intervals. The corresponding upper-tail ratios \(q_{0.95}/\mathrm{median}\), reported in Appendix Table~\ref{tab:waiting_time_summary}, confirm that return intervals have broad upper tails across the tested conditions. These curves quantify recurrence to previously visited sites and are distinct from the new-site discovery intervals \(\tau_n\), which measure the waiting time to discover previously unvisited sites.

Taken together, the short-biased step-length genomes, structured turning statistics, broad return-interval distributions, and consistently positive model-comparison scores support a common interpretation. In finite, depleting landscapes, the evolved search dynamics are better described as IS, combining local exploitation with occasional relocation, than as strict scale-free LW dynamics. Additional waiting-time analyses in Appendix~\ref{app:waiting_time_statistics} show that return and resource-hit intervals have broader upper tails than new-site discovery intervals, further supporting this interpretation.
\section{Discussion and conclusions}
We studied how search strategies evolve in finite, depletable resource landscapes when movement rules are inherited but not prescribed as a fixed parametric distribution. Across all grid sizes and resource environments, evolution shifted the step-length genomes from broadly distributed initial trait lists toward short-step-biased distributions with a sparse tail of longer entries. At the trajectory level, the resulting paths are consistent with local scanning interrupted by occasional relocation. Moment-based model comparison consistently favored IS over LW dynamics, with positive values of \(\Gamma\) across all environments and system sizes. Thus, within the present framework, adaptive search in finite depletion-driven landscapes is better described by IS than by strict scale-free LWs. The single-run sparse-density tests retained the IS fit preference. In a separate single-run ablation, agents were re-evolved using \(F=\eta_E\), and \(\Gamma\) remained positive in all six environments. In the fixed-genome cue-off evaluation, the mean score difference remained positive, but its magnitude changed in opposite directions across environments. These findings show that the coverage term is not solely responsible for the classification and that sensing affects the expressed trajectories. They do not establish cue-free evolution or isolate the effects of depletion and encounter-triggered stopping.

This result is consistent with the view that L\'evy-like search is not a universal optimum, but depends on ecological and physical constraints such as resource renewal, target density, memory, boundary effects, landscape heterogeneity, and sampling scale \cite{viswanathan1999optimizing,RevModPhys.87.483,10.1098/rsif.2014.1158,PhysRevResearch.6.023274,humphries2012foraging,10.1098/rsif.2008.0014,https://doi.org/10.1890/09-0079.1}. In the present model, resources are consumed upon encounter and do not regenerate during a lifetime. Under such conditions, a forager that has just encountered a resource may benefit from continued local search, because nearby resources are more likely to occur in clustered landscapes. However, local exploitation eventually becomes inefficient as the neighborhood is depleted. Occasional relocations are potentially useful because they move the agent away from over-explored regions. This interpretation is also consistent with movement-ecology work showing that animals in heterogeneous landscapes often alternate between intensive movement in profitable areas and more extensive movement between them \cite{Barraquand2008Heterogeneous}. The evolved strategy reflects this balance: short movements reduce cost and support local exploitation, while rarer longer movements reduce stagnation. The comparison with IS is also consistent with recent evolutionary work showing that, when the move-length distribution is not constrained to be a power law, evolved search can favor IS rather than strict L\'evy flight \cite{doi:10.1086/729220}. Our results extend this idea to explicitly depletable two-dimensional resource landscapes with genome-encoded step length, velocity, and turning traits. The evolved genomes do not produce a pure scale-free law. Instead, selection concentrates the encoded step-length pool at short values while preserving a small relocation component. This is closer to the structure expected for IS, where localized scanning and relocation are distinct functional components \cite{RevModPhys.83.81,10.1093/icb/41.2.137}, than to a strict L\'evy walk defined by a single heavy-tailed flight distribution.

The displacement-moment analysis provides a quantitative way to distinguish these alternatives. This is important because step-length or path-shape inspection alone can be sensitive to sampling choices and environmental structure, and can therefore blur the distinction between intrinsic movement rules and environment-induced movement patterns \cite{10.1098/rsif.2008.0014,https://doi.org/10.1890/09-0079.1,Barraquand2008Heterogeneous}. The second and fourth moments probe different aspects of the trajectory: the second moment captures the typical spread of displacement with lag, whereas the fourth moment is more sensitive to rare large displacements. Across environments, the IS model gives a closer moment-level description, and the preference remains positive across the three grid sizes. 
This supports the preference for the IS description in this two-model comparison, without identifying a unique generating mechanism or establishing superior foraging performance. The result is also consistent with the moment-based distinction between L\'evy and IS developed in Refs.~\cite{2bzm-t9k1,BHANDARI2025102334}. The waiting-time statistics provide a complementary view of the same dynamics. The running maximum of return intervals shows that extreme returns to already visited sites arise as rare record events, while the appendix analysis separates recurrence, resource encounter timing, and new-site discovery. Return intervals and resource-hit intervals have broader upper tails than new-site discovery intervals, indicating that the strongest temporal intermittency appears in recurrence and resource-encounter timing rather than in long pauses in spatial discovery. This distinction is important because the visitation-statistics quantity \(\tau_n\) refers to the time needed to discover a new site after \(n\) distinct sites have already been visited \cite{PhysRevLett.132.127101}. In our analysis, \(\tau_n\) is therefore treated separately from return intervals to previously visited sites.

These findings connect evolutionary foraging with random-walk models of depletion-controlled search. Starving random walks and related visitation-statistics models emphasize that survival depends not only on the total number of sites visited, but also on the temporal structure of new-site discovery \cite{ben-Avraham_Havlin_2000,Havlin01011987,PhysRevLett.132.127101}. Our simulations do not implement starvation through a hard metabolic deadline, but they share the same depletion principle: continued success requires movement into regions where resources remain available. The evolved strategies therefore occupy an intermediate position between idealized random-walk theory and adaptive foraging. They are shaped by local resource geometry, depletion, movement cost, and heritable variation in movement traits. The results also suggest a possible design principle for resource-limited search and optimization. Movement rules dominated by short displacements but punctuated by occasional relocations may be useful in heterogeneous objective landscapes, especially when local improvement is likely but prolonged exploitation becomes inefficient \cite{RevModPhys.87.483,doi:10.5772/60414,leon2021mutations}. In autonomous search, a comparable rule would combine local scanning after detection with occasional relocation to avoid spending excessive time in depleted or low-yield regions. The reactive stopping rule in our simulations, in which movement segments terminate upon resource encounter, is a simple instance of this principle and may be relevant to energy-limited robotic or algorithmic exploration \cite{viswanathan2011physics}.

Several features of the model limit the interpretation of these results. The search domain is finite and periodic, so the simulations do not access the asymptotic regimes assumed in many analytical LW or random-walk results. Resource landscapes are fixed during a lifetime except for depletion, and no resource renewal is included. The evolutionary protocol also uses fixed population size, mutation parameters, and genome length across all environments. Independent replication is limited to \(L=503\) and \(\rho=0.15\). The other grid sizes and density conditions remain single-run tests. Finally, the simulations are single-agent evaluations and do not include social information, competition, or collective effects, which can alter the balance between exploration and exploitation \cite{10.1177/26339137241228858,pyke2022reinforcement,Garg2022CollectiveForaging}. Future work should examine larger domains, explicit resource renewal, and alternative sensing rules to determine when intermittent strategies persist and when scale-free movement becomes advantageous. Multi-agent extensions would also be useful for studying how cooperation, interference, or social learning modifies the evolved search strategy. More broadly, although the present work is a controlled simulation study rather than a deployable system, search rules of this kind could support energy-aware autonomous exploration and ecological monitoring. Still, they would require privacy, safety, and oversight safeguards if used in sensitive human or environmental settings. In summary, evolution on finite, depletable landscapes drives movement genomes toward short-step-biased trait distributions with occasional relocation, and the IS dynamics more accurately describe the resulting trajectories than strict LW within the moment-based classification used here. At \(L=503\) and \(\rho=0.15\), independent runs support a reproducible IS--LW fit preference, conditional on the model's sensing, depletion, and encounter-stopping rules.

\section*{Acknowledgments}
We thank Pedro Lencastre for providing synthetic trajectories used to validate our moment-based classifier.
The authors 
acknowledge funding from the Research Council of Norway under grant number 335940 for the project \emph{Virtual-Eye}. The work of AY was also supported by the Center of Excellence, \emph{Integreat}, project number 332645.

\bibliographystyle{unsrt}
\bibliography{references}
\appendix
\section{The evolutionary algorithm: specifications}
\label{app:evolutionary_algorithm}
We use a generational genetic algorithm to evolve a population of \(N_{\mathrm{pop}}\) haploid agents over \(n_{\mathrm{gen}}\) generations. Each agent carries three independent integer-valued genomes of fixed length \(G\), corresponding to step length, velocity, and turning-angle preference:
\begin{align*}
\mathbf g^{(s)} &\in \{1,2,\dots,\lfloor L/2\rfloor\}^G,
&& \text{step-length genome}, \\
\mathbf g^{(v)} &\in \{1,2,\dots,v_{\max}\}^G,
&& \text{velocity genome}, \\
\mathbf g^{(\theta)} &\in \{0,1,\dots,n_\theta-1\}^G,
&& \text{turn-angle genome}.
\end{align*}
Here, \(L\) is the linear grid size, \(v_{\max}=L\) is the maximum admissible velocity, and \(n_\theta=36\) is the number of turning-angle bins, corresponding to angular increments $\theta_\kappa = 10^\circ \kappa$, $\kappa=0,1,\dots,35$. The genomes are represented as finite lists of admissible trait values rather than as explicit probability distributions. During trajectory generation, phenotypic decisions are obtained by sampling uniformly from these lists. Evolution, therefore, acts on the empirical frequencies of encoded trait values, allowing the movement statistics to emerge from selection, recombination, and mutation.

Each lifetime is simulated using a fresh copy of the initial pre-depletion resource map associated with that evolutionary run, so resource depletion is reset between agent evaluations. At the beginning of each lifetime, an agent is initialized at a resource-bearing lattice site chosen uniformly at random. If no resource-bearing site is available, the initial position is drawn uniformly from the lattice. Heading selection follows the complete rule in Sec.~\ref{sec:agent}. With probability $(p_{\mathrm{cue}})$, the cue branch searches the clipped square sensing window. If the window contains an undepleted resource, the heading is set directly toward the nearest resource by Euclidean distance. If the window contains no undepleted resource, a uniformly random heading is selected without returning to the genome branch. With probability $(1-p_{\mathrm{cue}})$, the genome-sampled turning increment is applied. Only the lattice position at the end of each movement update is checked for resources. If an undepleted resource is present at that endpoint, it is consumed, the harvested energy is incremented, and the current movement segment terminates at that endpoint. Table~\ref{tab:simulation_parameters} lists the set of simulation and evolutionary parameters used in the reported experiments.
\subsection*{Fitness evaluation}\label{sec:fitness_appendix}
Fitness is evaluated using the definition given in Sec.~\ref{sec:fitness}. For each agent, the quantities entering Eq.~\eqref{eq:fitness_expanded} are accumulated along the realized trajectory, and the terminal value of $\mathcal F$ is used as the reproductive score. Harvested energy is updated whenever a resource is encountered and consumed. Movement cost is accumulated from the scalar movement increment $(\Delta_j)$. Distinct lattice sites are tracked explicitly, and each return to an already visited site increments the revisit count. Thus, fitness is assigned at the end of the lifetime of each individual.

\begin{figure}[t]
\centering
\includegraphics[width=\textwidth]{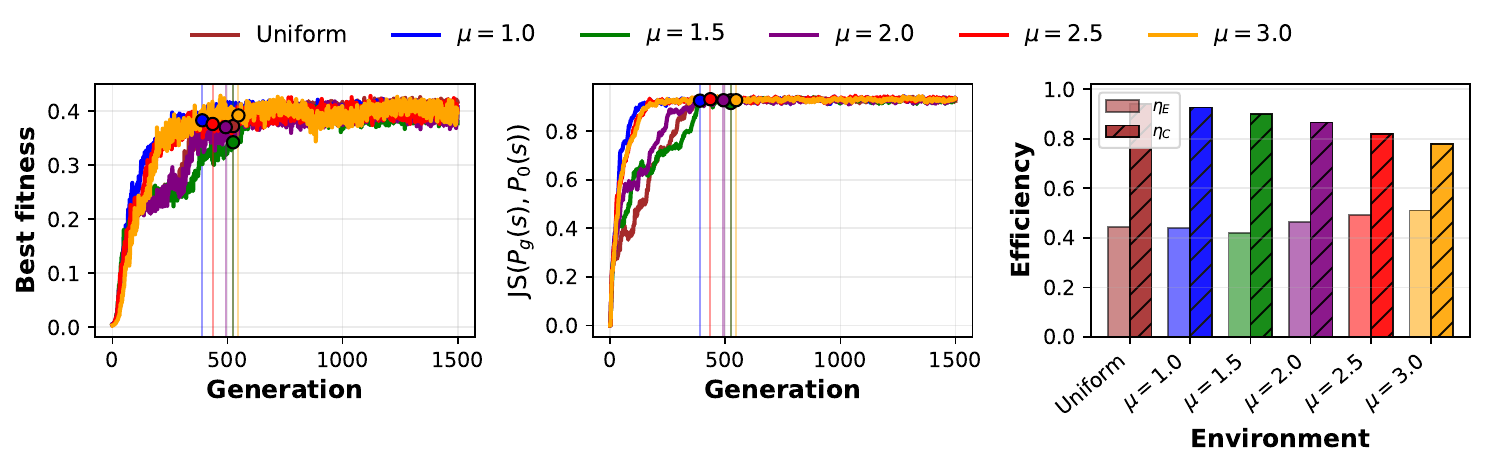}
\caption{
Evolutionary stabilization on the \(503\times503\) grid.
Left: best fitness as a function of generation.
Middle: JS divergence \(\mathrm{JS}(P_g(s),P_0(s))\) between the population-level step-length distribution at generation \(g\) and the initial distribution.
Right: final energetic and coverage efficiencies, \(\eta_E\) and \(\eta_C\), for the best evolved agent in each environment.
Vertical markers denote the operational plateau generation estimated from a rolling-slope criterion applied to fitness and genome redistribution.
}
\label{fig:fitness_decomposition_js_appendix}
\end{figure}

Figure~\ref{fig:fitness_decomposition_js_appendix} summarizes three aspects of the evolutionary dynamics on the \(503\times503\) grid. The left panel shows that the best fitness rises rapidly during the early generations and then fluctuates around an environment-dependent plateau. Because mutation and recombination remain active throughout evolution, this plateau should be interpreted as an operational stabilization regime rather than as exact convergence.
\begin{table}[t]
\centering
\caption{Simulation and evolutionary parameters used for the main experiments. All main results use resource density \(\rho=0.15\). The additional \(503\times503\) density robustness analysis uses
\protect\(\rho\in\{0.005,0.010,0.30,0.45\}\)}.
\label{tab:simulation_parameters}
\renewcommand{\arraystretch}{1.15}
\setlength{\tabcolsep}{6pt}
\begin{tabular}{p{0.35\linewidth}p{0.58\linewidth}}
\toprule
Parameter & Value \\
\midrule
Grid sizes \(L\) & \(211,\;503,\;1009\) \\
Environment types & uniform,\; \(\mu=1.0,\;1.5,\;2.0,\;2.5,\;3.0\) \\
Resource density \(\rho\) & \(0.15\) for all grid sizes, with \(0.005,\;0.010,\;0.30,\;0.45\) additionally tested at \(L=503\) \\
Resource value & \(1.0\) \\
Genome length \(G\) & \(30\) \\
Population size \(N_{\mathrm{pop}}\) & \(40\) \\
Number of generations \(n_{\mathrm{gen}}\) & \(1500\) \\
Step-length range & \(1,\dots,\lfloor L/2\rfloor\) \\
Velocity range & \(1,\dots,L\) \\
Turn-angle bins & \(0^\circ,10^\circ,\dots,350^\circ\) \\
Sensory cue & \(p_{\mathrm{cue}}=0.5,\;r_{\mathrm{cue}}=60\), using a square window clipped \\
Lifespan \(T_{\mathrm{life}}\) & \(8000\) \\
Movement cost coefficient \(c_{\mathrm{move}}\) & \(0.5\) \\
Mutation count per genome \(n_{\mathrm{mut}}\) & \(6\) loci \\
Number of angle bins \(n_\theta\) & \(36\) \\
Resource gain \(\varepsilon\) & \(1.0\) \\
Elitism fraction \(e_{\mathrm{elite}}\) & \(0.1\) \\
Elite count \(N_{\mathrm{elite}}\) & \(\lfloor e_{\mathrm{elite}}N_{\mathrm{pop}}\rfloor\) \\
Mutation scale \(\sigma_{\mathrm{mut}}\) & \(2\) \\
Randomization
& seed \(42\) for the single-run conditions, with independent resource maps and GA initializations for the five replicated \(L=503,\rho=0.15\) runs \\
Movement boundary condition & periodic \\
\bottomrule
\end{tabular}
\end{table}

The middle panel shows the cumulative Jensen--Shannon divergence \cite{MENENDEZ1997307} \(\mathrm{JS}(P_g(s),P_0(s))\), where \(P_g(s)\) is the normalized population-level distribution of genome-encoded step lengths at generation \(g\). The divergence rises rapidly and then saturates, indicating that selection reorganizes the step-length genome early in evolution and subsequently maintains a comparatively stable evolved distribution. To quantify stabilization, we used a rolling-slope criterion. For each environment, rolling linear slopes were computed over a window of 100 generations for both best fitness and \(\mathrm{JS}(P_g(s),P_0(s))\). The plateau generation was defined as the first generation after \(g=150\) for which both absolute slopes remained below fixed thresholds for 50 consecutive generations. With thresholds \(2\times 10^{-4}\) for fitness and \(5\times 10^{-4}\) for JS divergence, the estimated plateau generations ranged from \(g=390\) to \(g=547\) across environments. These values support the interpretation of rapid early adaptation followed by stabilization, without implying exact convergence.

The right panel decomposes the final best-agent fitness into its energetic and coverage components, \(\eta_E\) and \(\eta_C\). Since \(\mathcal F=\eta_E\eta_C\), these values show how final performance is assembled from energy acquisition relative to movement cost and from low-redundancy spatial coverage. Across environments, \(\eta_C\) remains high but decreases from \(0.94\) in the uniform case to \(0.78\) at \(\mu=3.0\), whereas \(\eta_E\) ranges from \(0.42\) to \(0.51\). Thus, final fitness reflects both energetic cost and the increasing burden of revisits in some environments.
\subsection*{Selection, recombination, and mutation}
\label{sec:selection_appendix}
\begin{algorithm}[t]
\caption{Generational evolutionary optimization of foraging strategies}
\label{alg:evolutionary_foraging}
\begin{algorithmic}[1]
\REQUIRE Population size \(N_{\mathrm{pop}}\), number of generations \(n_{\mathrm{gen}}\), genome length \(G\), elitism fraction \(e_{\mathrm{elite}}\), mutation count \(n_{\mathrm{mut}}\), mutation scale \(\sigma_{\mathrm{mut}}\)
\ENSURE Final evolved population \(\mathcal{P}^{(n_{\mathrm{gen}})}\)
\STATE Randomly initialize the population \(\mathcal{P}^{(0)}=\{X_i^{(0)}\}_{i=1}^{N_{\mathrm{pop}}}\)
\FOR{\(g=0\) to \(n_{\mathrm{gen}}-1\)}
    \FOR{each individual \(X_i^{(g)} \in \mathcal{P}^{(g)}\)}
        \STATE Simulate its lifetime trajectory in the current environment
        \STATE Compute fitness \(\mathcal{F}_i^{(g)}\) according to Eq.~\eqref{eq:fitness_expanded}
    \ENDFOR
    \STATE Rank all individuals by fitness
    \STATE Copy the top \(N_{\mathrm{elite}}=\lfloor e_{\mathrm{elite}}N_{\mathrm{pop}} \rfloor\) individuals into \(\mathcal{E}^{(g)}\)
    \STATE Set parent-selection probabilities \(p_i\) according to Eq.~\eqref{eq:selection_probability_appendix}
    \STATE Initialize the offspring set \(\mathcal{O}^{(g)}=\varnothing\)
    \WHILE{\(|\mathcal{E}^{(g)}|+|\mathcal{O}^{(g)}|<N_{\mathrm{pop}}\)}
        \STATE Sample two parents from \(\mathcal{P}^{(g)}\) using probabilities \(\{p_i\}\)
        \STATE Generate offspring by independent trait-wise recombination as in Eq.~\eqref{eq:recombination_appendix}
        \STATE Apply bounded Gaussian integer mutation as in Eq.~\eqref{eq:mutation_operator_appendix}
        \STATE Add the resulting offspring to \(\mathcal{O}^{(g)}\)
    \ENDWHILE
    \STATE Form the next generation:$\mathcal{P}^{(g+1)}
    =\mathcal{E}^{(g)}\cup
    \mathcal{O}^{(g)}_{[1:N_{\mathrm{pop}}-N_{\mathrm{elite}}]}$
\ENDFOR
\STATE \textbf{return} \(\mathcal{P}^{(n_{\mathrm{gen}})}\)
\end{algorithmic}
\end{algorithm}
At each generation, all individuals are evaluated independently in the current environment and ranked by fitness. The top fraction \(e_{\mathrm{elite}}\) is copied unchanged into the next generation, implementing elitism. In the reported experiments, \(e_{\mathrm{elite}}=0.1\), so the elite count is \(N_{\mathrm{elite}}=\lfloor e_{\mathrm{elite}}N_{\mathrm{pop}}\rfloor\).

The remaining individuals are generated from parents sampled with probability proportional to fitness. If \(\mathcal F_i\) denotes the fitness of individual \(i\), then the selection probability is
\begin{equation}
p_i= \frac{\max(\mathcal F_i,\varepsilon_s)}
{\sum_{j=1}^{N_{\mathrm{pop}}}\max(\mathcal F_j,\varepsilon_s)},
\label{eq:selection_probability_appendix}
\end{equation}
where \(\varepsilon_s>0\) is a small numerical constant introduced to avoid vanishing probabilities when fitness values are extremely small.

Recombination is applied independently to each genome type. Because the genomes are interpreted as unordered lists of admissible trait values rather than ordered chromosomal sequences, each parental genome is first randomly permuted before crossover. Let \(\mathbf a\) and \(\mathbf b\) denote two parental genomes of the same type, each of length \(G\), and let \(\pi_a\) and \(\pi_b\) denote independent random permutations of \(\{1,\dots,G\}\). Writing $\mathbf a^{\pi_a}=(a^{\pi_a}_1,\dots,a^{\pi_a}_G),\;\;
\mathbf b^{\pi_b}=(b^{\pi_b}_1,\dots,b^{\pi_b}_G)$, the offspring genomes are formed by exchanging the second half after permutation:
\begin{equation}
\mathbf c_1=
\bigl(
a^{\pi_a}_1,\dots,a^{\pi_a}_m,
b^{\pi_b}_{m+1},\dots,b^{\pi_b}_G
\bigr),
\qquad
\mathbf c_2=
\bigl(
b^{\pi_b}_1,\dots,b^{\pi_b}_m,
a^{\pi_a}_{m+1},\dots,a^{\pi_a}_G
\bigr),
\label{eq:recombination_appendix}
\end{equation}
where \(m=\lfloor G/2\rfloor\). This operator mixes parental trait frequencies while avoiding an artificial dependence on locus order.

Mutation is then applied independently to each of the three genomes. In the reported experiments, exactly \(n_{\mathrm{mut}}=6\) loci were selected uniformly at random within each genome of length \(G=30\). If \(a_i\) is the value at a selected locus, the mutated value is
\begin{equation}
a_i'=
\operatorname{clip}\!\left(
a_i+\operatorname{round}(Z),
\,a_{\min},\,a_{\max}
\right),
\qquad
Z\sim\mathcal N(0,\sigma_{\mathrm{mut}}^2),
\label{eq:mutation_operator_appendix}
\end{equation}
with \(\sigma_{\mathrm{mut}}=2\). The clipping operation enforces the admissible bounds of the corresponding genome: \(1\le s\le \lfloor L/2\rfloor\) for move lengths, \(1\le v\le L\) for velocities, and \(0\le \kappa\le 35\) for turn-angle indices. The mutation and recombination hyperparameters were fixed after pilot exploration and then kept identical across all environments and grid sizes. The aim was not to optimize the evolutionary algorithm separately for each condition, but to use a single moderate setting that preserves heritability while maintaining sufficient exploratory variation. In particular, mutating 6 of 30 loci per genome provides a moderate perturbation rate, while the Gaussian mutation scale \(\sigma_{\mathrm{mut}}=2\) yields predominantly local integer changes. Likewise, shuffled half-and-half recombination was chosen because the genomes encode unordered trait-value lists rather than ordered genetic sequences. Keeping these choices fixed across all conditions reduces the risk that the reported differences between environments arise from retuning the evolutionary operators rather than from the ecological structure of the landscapes.

The full evolutionary cycle is summarized in Algorithm~\ref{alg:evolutionary_foraging}. At each generation, fitness is first evaluated for the entire population. Elite individuals are copied directly into the next generation, while the remainder of the population is filled by offspring generated through fitness-proportionate parent selection, recombination, and mutation. This procedure is repeated for the prescribed number of generations. Fitness evaluation is parallelized across agents, since trajectories are conditionally independent once the environment and the genomes are specified. In addition to the final population, the implementation stores generation-wise fitness values, summary statistics, and genome histories, enabling exact post hoc analysis of the evolutionary dynamics.

\subsection*{Implementation and compute resources}

All simulations were executed in a Python 3.11 environment using a Docker-based GitLab CI runner. Fitness evaluation was parallelized on the CPU using a \texttt{ProcessPoolExecutor} with up to 8 worker processes. The execution host had an NVIDIA RTX A6000 GPU available (49\,140, MiB VRAM; driver version 580.65.06, CUDA 13.0), although the simulation code itself did not use GPU-specific acceleration. A full experiment suite, comprising one complete run across all six environments for 1500 generations, required approximately 6-10 hours of wall-clock time.

\section{Additional robustness results}
\label{app:robustness_results}
\subsection*{Results across different grid sizes}
\label{app:grid_sizes}
\begin{figure}[ht!] 
\centering
\fbox{\includegraphics[width=0.44\textwidth]{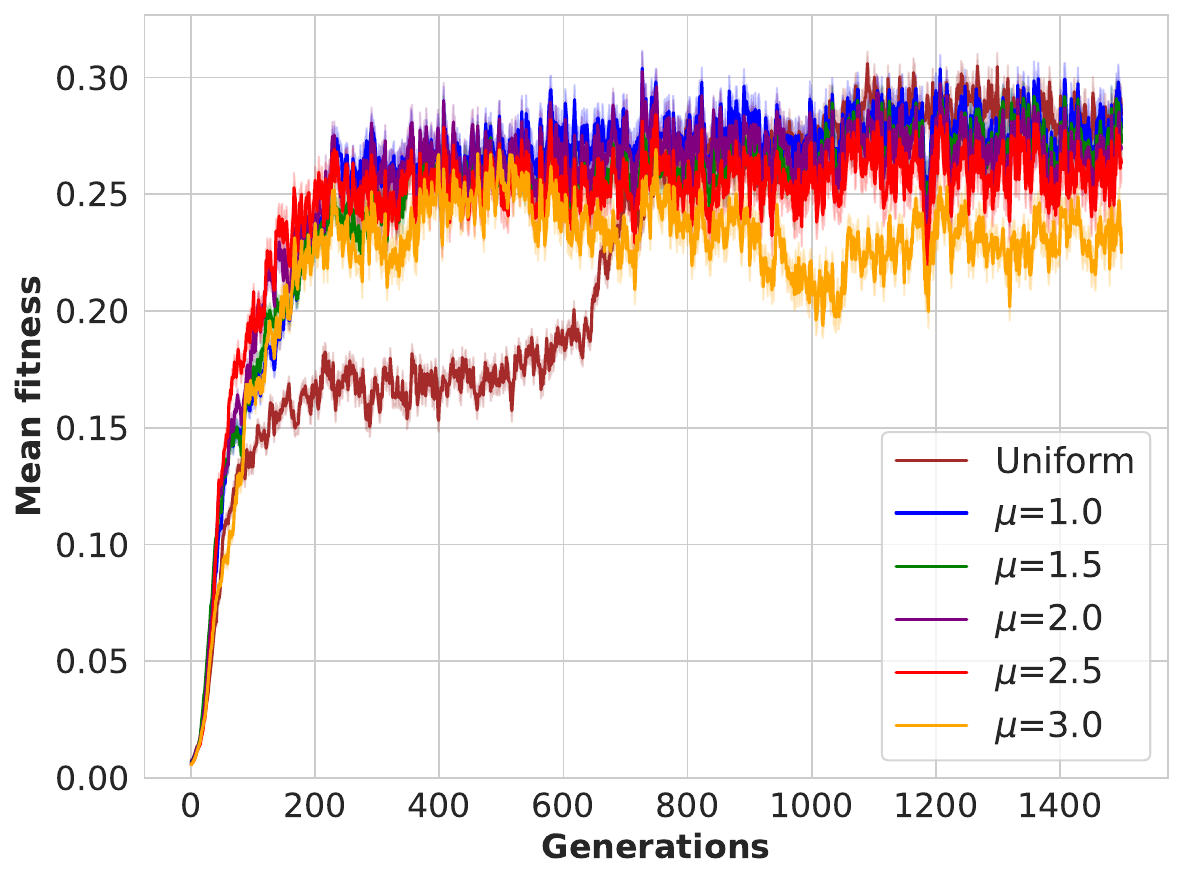}\hfill
\includegraphics[width=0.54\textwidth]{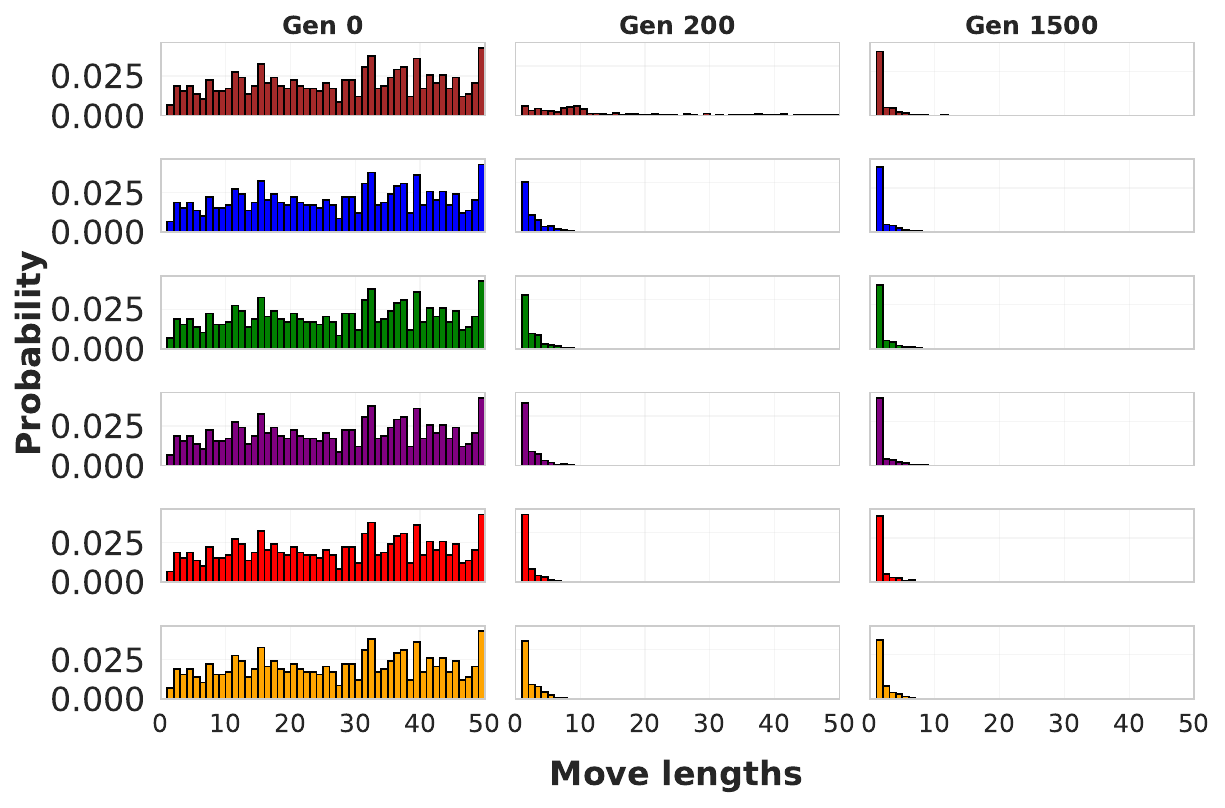}}
\fbox{\includegraphics[width=0.98\textwidth]{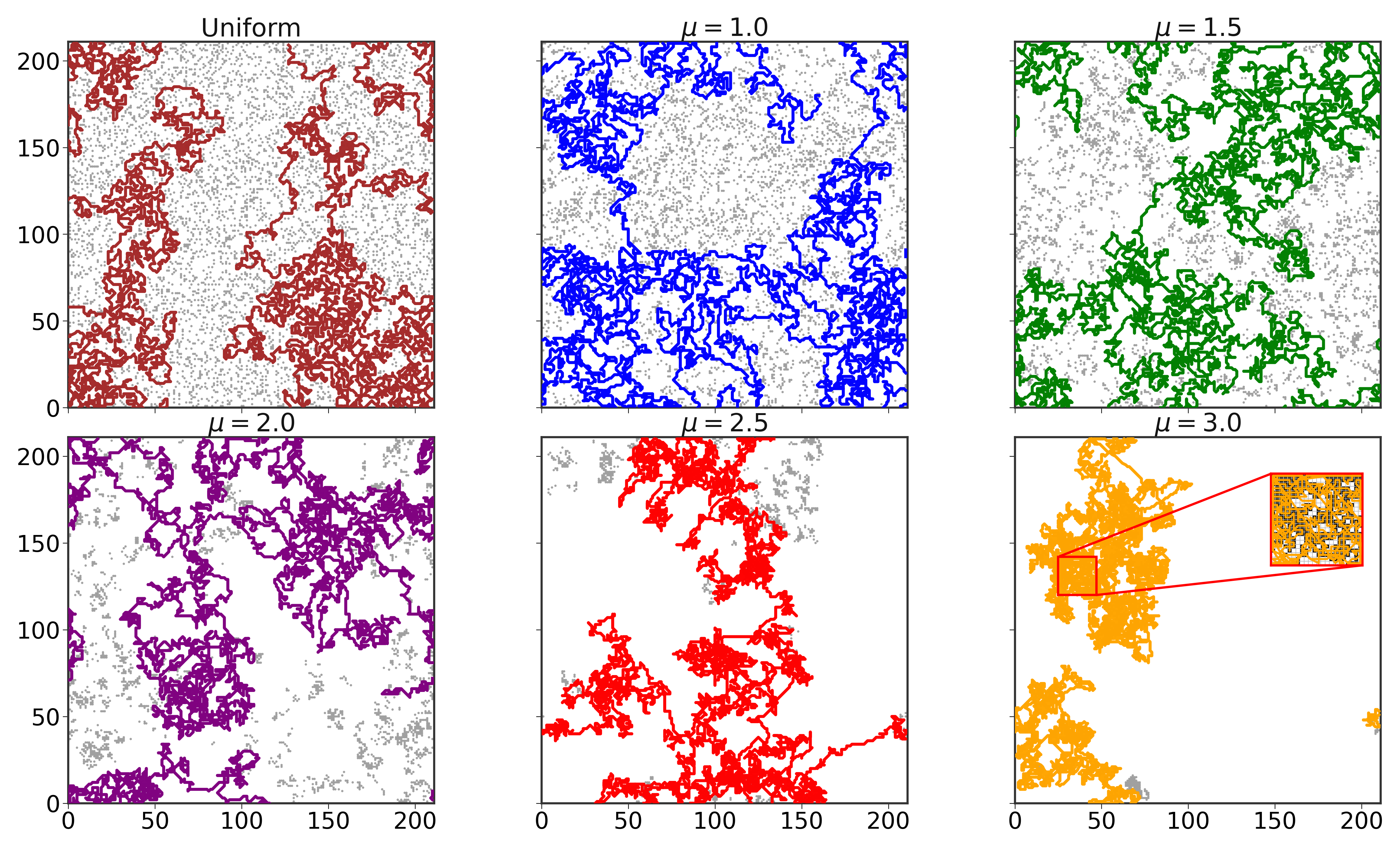}}
\fbox{\includegraphics[width=0.98\textwidth]{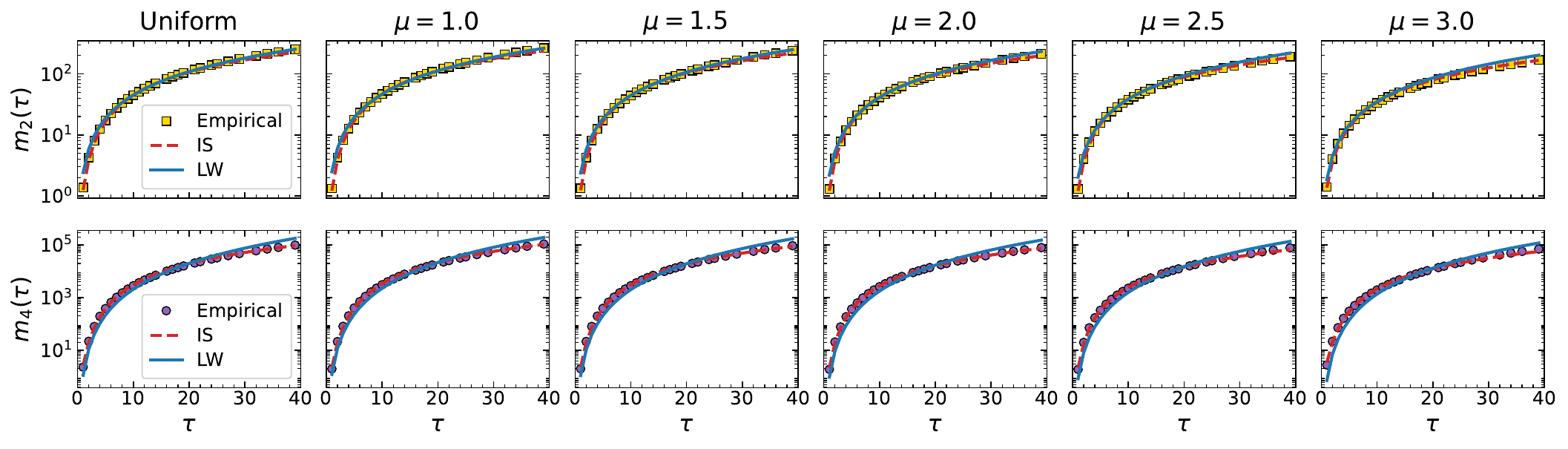}}
\caption{
Additional-size results for the \(211\times211\) torus.
\textbf{Top left}: mean fitness over 1500 generations for each resource environment, with shaded standard errors  across agents within the single evolutionary run at each generation.
\textbf{Top right}: distributions of genome-encoded step-length entries at generations 0, 200, and 1500.
\textbf{Middle}: resource maps overlaid with the trajectory of the best evolved agent in each environment.
\textbf{Bottom}: empirical second and fourth displacement moments compared with best-fit IS and LW models.
}
\label{fig:evolution211x211_appex}
\end{figure}

\begin{figure}[ht!] 
\centering
\fbox{\includegraphics[width=0.44\textwidth]{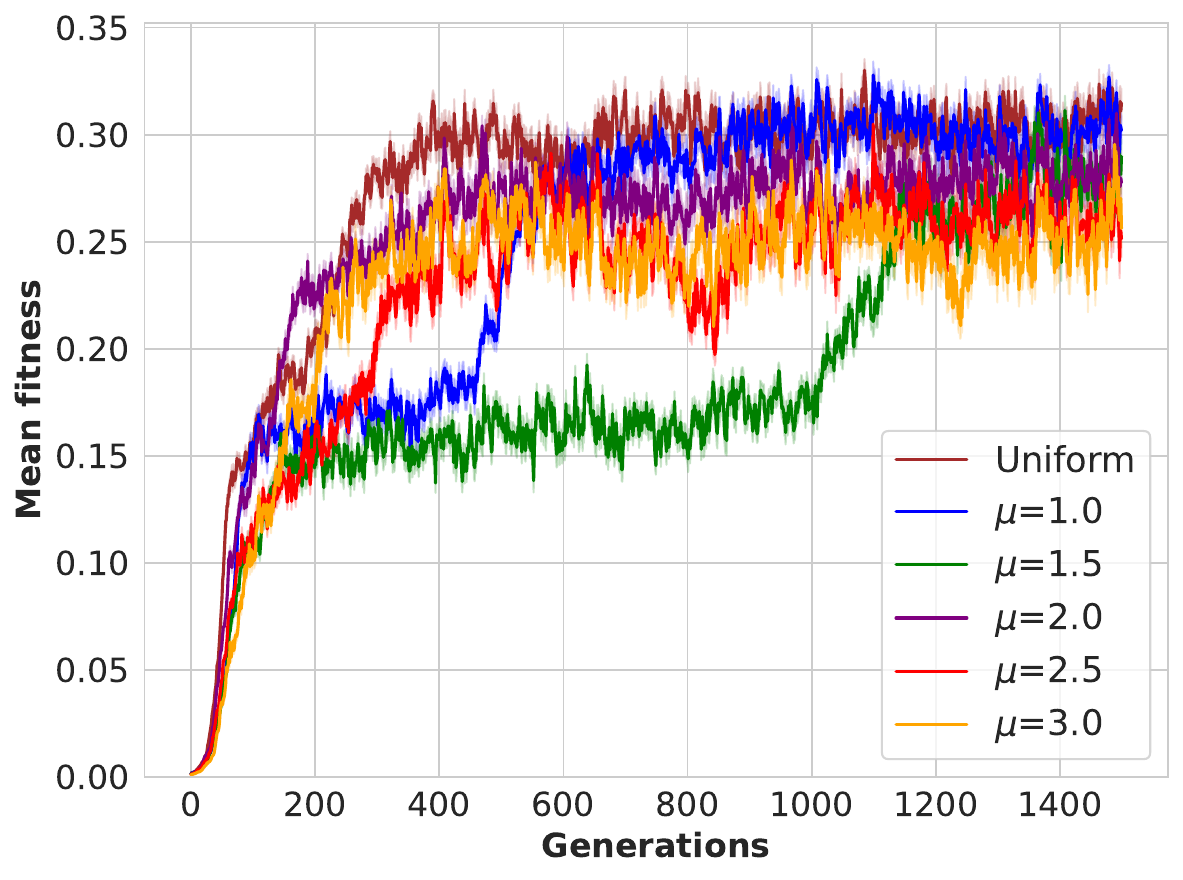}\hfill
\includegraphics[width=0.54\textwidth]{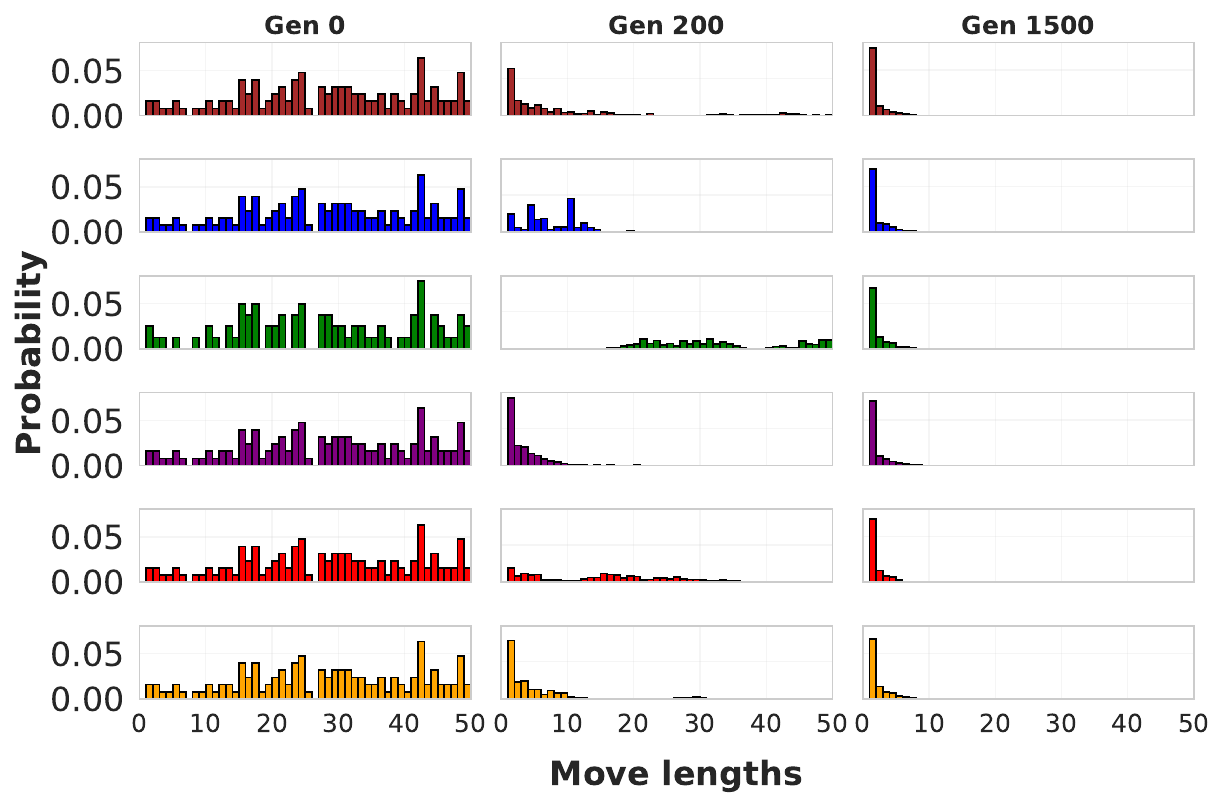}}
\fbox{\includegraphics[width=0.98\textwidth]{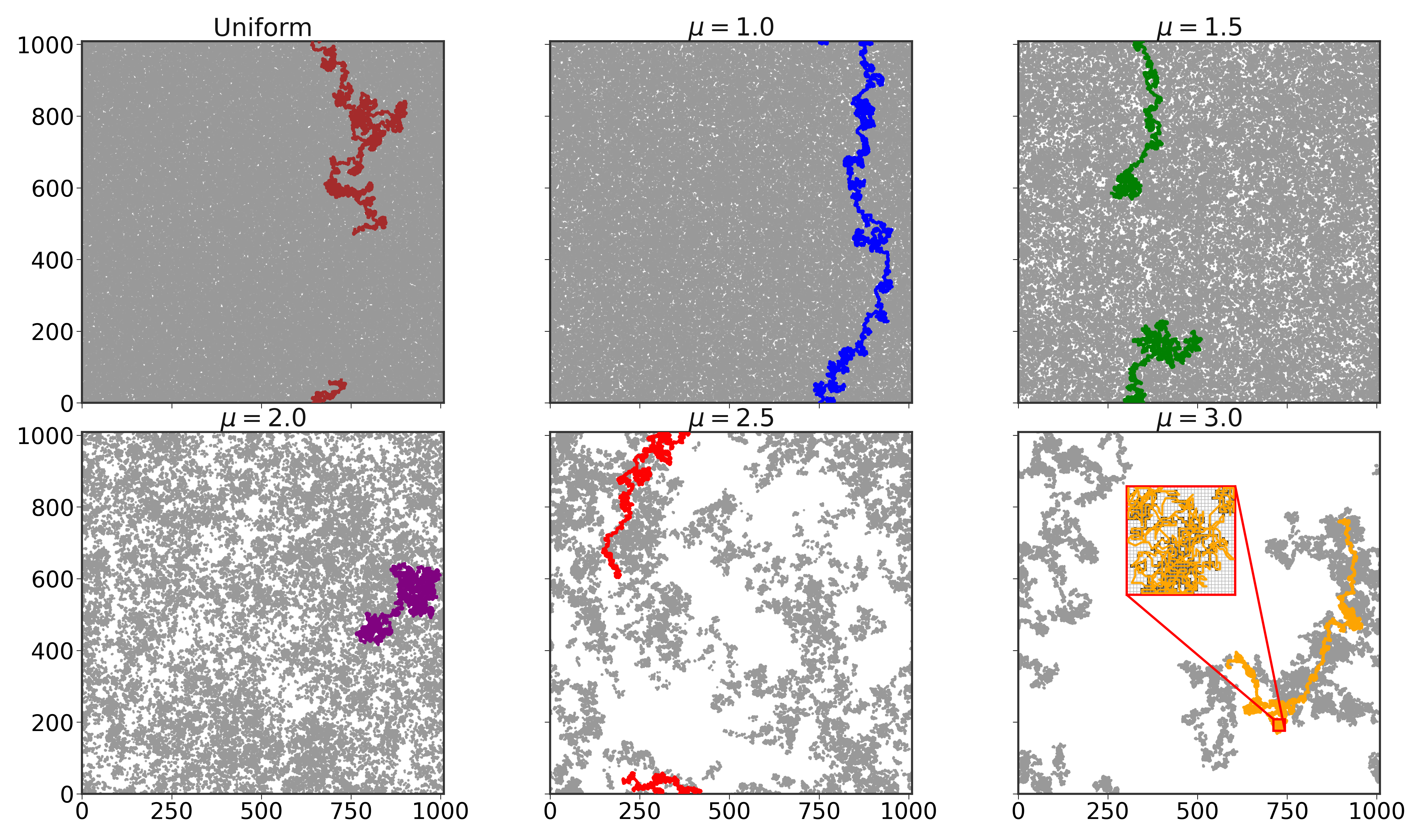}}
\fbox{\includegraphics[width=0.98\textwidth]{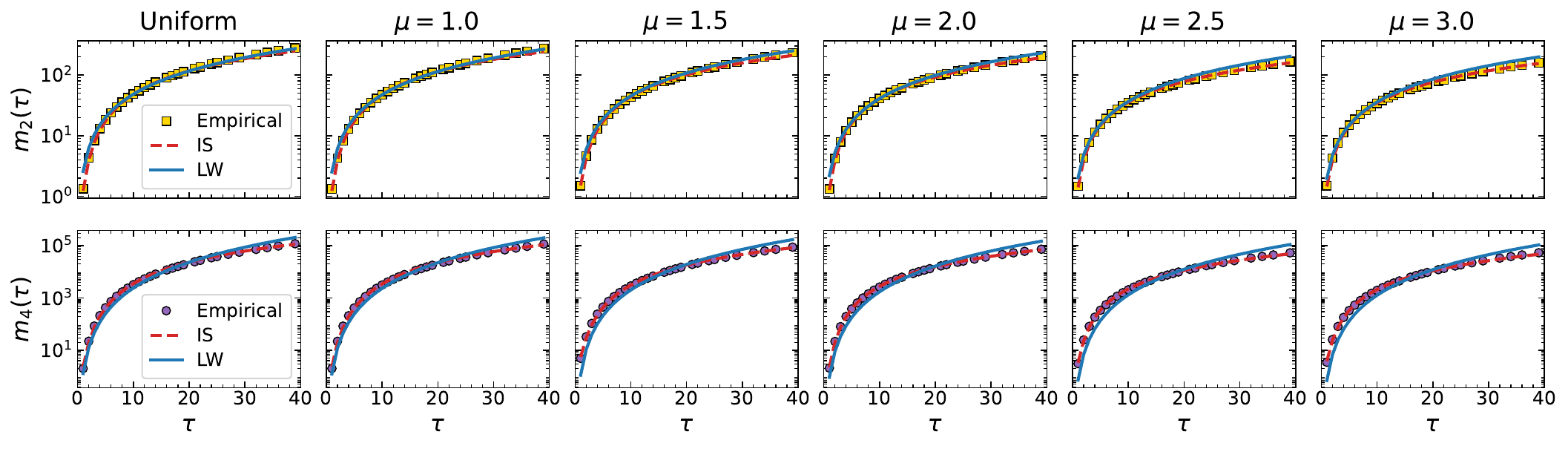}}
\caption{Additional-size results for the \(1009\times1009\) torus. \textbf{Top left}: mean fitness over 1500 generations for each resource environment, with shaded standard errors across agents within the single evolutionary run at each generation. \textbf{Top right}: distributions of genome-encoded step-length entries at generations 0, 200, and 1500.
\textbf{Middle}: resource maps overlaid with the trajectory of the best evolved agent in each environment.
\textbf{Bottom}: empirical second and fourth displacement moments compared with best-fit IS and LW models.} 
\label{fig:evolution_appex}
\end{figure}

We tested whether the main results depend on the linear system size by repeating the analysis on \(211\times211\) and \(1009\times1009\) tori. These additional domains differ in both spatial extent and total resource capacity, while preserving the same resource classes, evolutionary protocol, and fitness definition used for the \(503\times503\) system. Figures~\ref{fig:evolution211x211_appex} and \ref{fig:evolution_appex} show the corresponding fitness trajectories, genome-encoded step-length distributions, best evolved trajectories, and displacement-moment fits. In both additional system sizes, mean fitness increases during the early generations and then fluctuates around an environment-dependent plateau. The step-length genomes show the same qualitative evolutionary change as in the main system. Initially, broad trait lists become concentrated at short values while retaining a sparse tail of longer entries.

The evolved trajectories also preserve the same broad movement organization across system sizes. They combine local scanning with occasional relocation, although the detailed path geometry varies with resource environment and domain size. Larger domains allow longer realized transfers, while smaller domains impose stronger spatial constraints through the finite torus. These differences affect quantitative details but do not change the main qualitative pattern.

The bottom panels of Figures~\ref{fig:evolution211x211_appex} and \ref{fig:evolution_appex} compare empirical second and fourth displacement moments with the corresponding IS and LW fits. The same model-selection pattern observed on the \(503\times503\) grid persists at both additional sizes. Across all environments and grid sizes, the score difference \(\Gamma=\bar R^2_{\mathrm{IS}}-\bar R^2_{\mathrm{LW}}\) remains positive, as reported in Table~\ref{tab:model_comparison_summary}. Thus, the preference for IS dynamics is not restricted to a single spatial scale.

\subsection*{Results across different resource densities}
\label{app:density_results}
\begin{figure}[t]
\centering
\includegraphics[width=\textwidth]{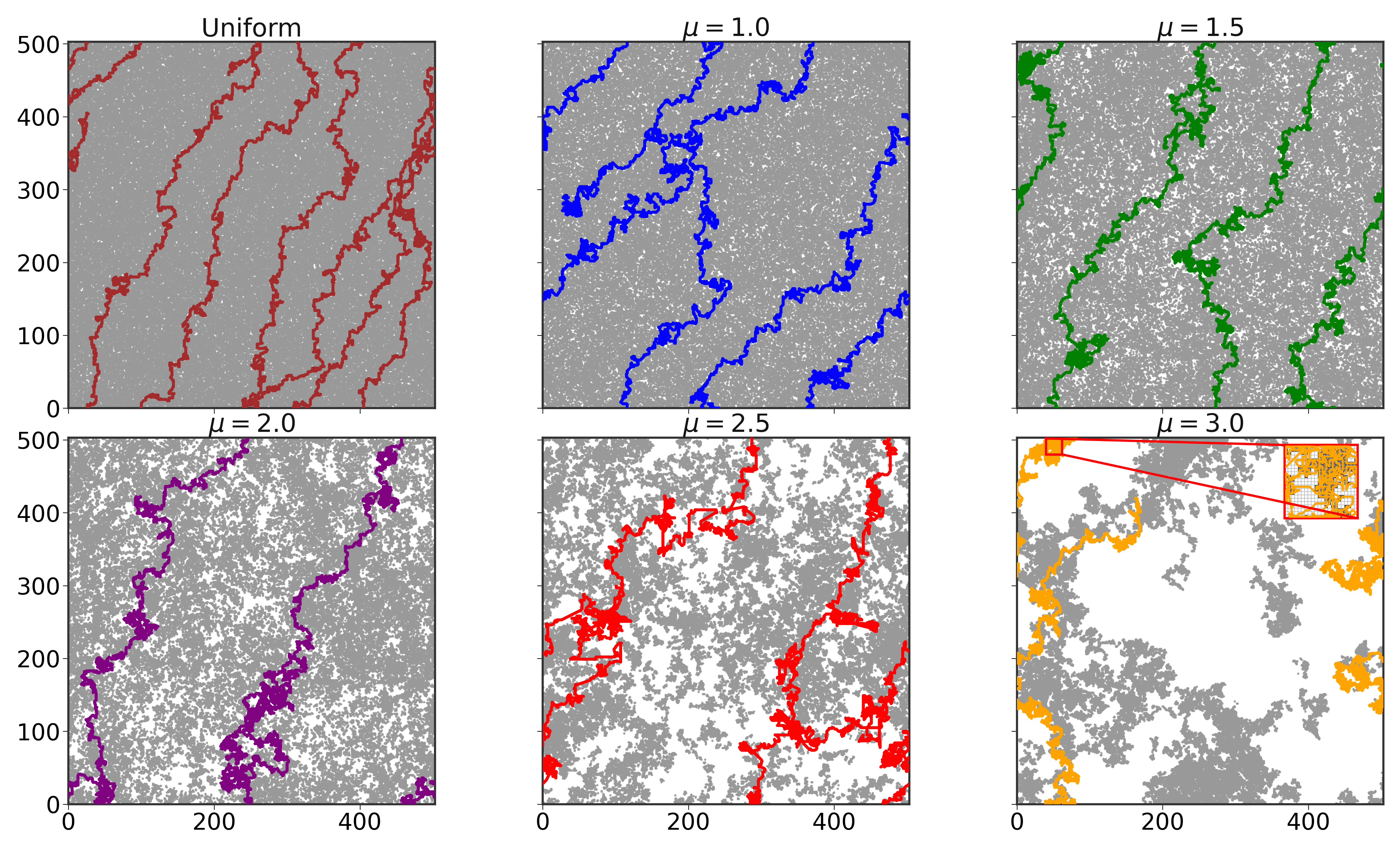}
\includegraphics[width=0.56\textwidth]{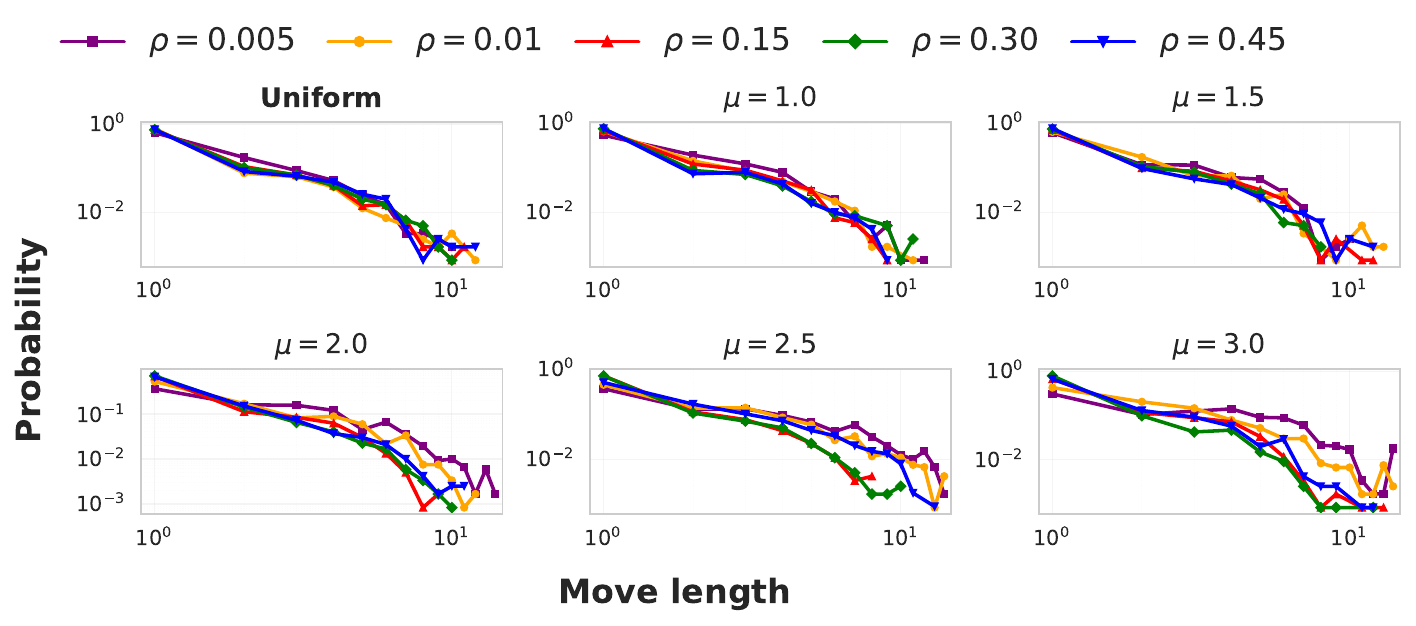}
\hfill
\includegraphics[width=0.43\textwidth]{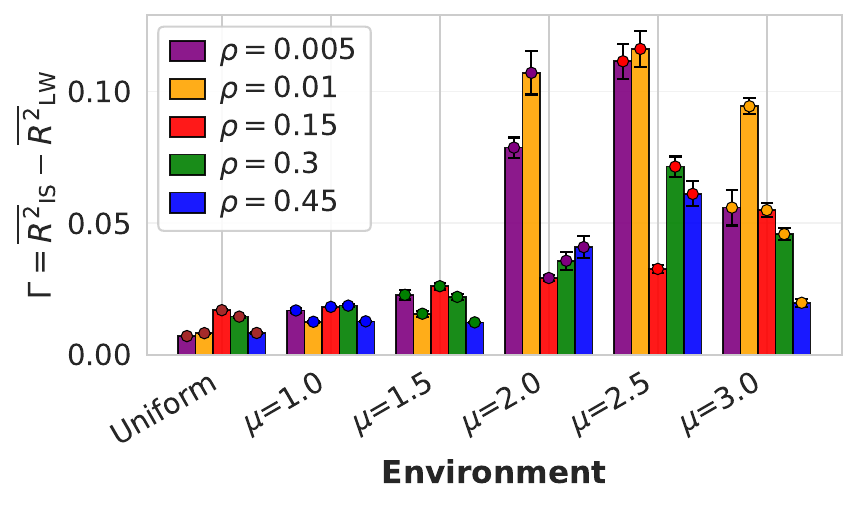}
\caption{\protect
Robustness of evolved search dynamics across resource densities on the \(503\times503\) torus.
\textbf{Top}: best evolved trajectories overlaid on the corresponding resource maps for the highest tested density, \textbf{\(\rho=0.45\)}. \textbf{Bottom left}: final-population distributions of genome-encoded step lengths for \(\rho\in\{0.005,0.010,0.15,0.30,0.45\}\), shown separately for each resource environment. \textbf{Bottom right}: model-selection score \(\Gamma\) across densities, with error bars.}
\label{fig:density_results_503}
\end{figure}

\begin{table}[t]
    \centering
    \small
    \caption{Realized resource occupancy $\rho_{\mathrm{real}}$, the fraction of distinct resource-bearing sites before depletion, for the saved $503\times503$ maps at the three original nominal densities. Each entry describes the corresponding resource map.}
    \label{tab:realized_occupancy}
    \begin{tabular}{lccc}
        \hline
        Environment & $\rho=0.15$ & $\rho=0.30$ & $\rho=0.45$ \\
        \hline
        Uniform   & 0.150 & 0.299 & 0.449 \\
        $\mu=1.0$ & 0.138 & 0.256 & 0.359 \\
        $\mu=1.5$ & 0.131 & 0.245 & 0.344 \\
        $\mu=2.0$ & 0.117 & 0.221 & 0.310 \\
        $\mu=2.5$ & 0.097 & 0.184 & 0.263 \\
        $\mu=3.0$ & 0.066 & 0.145 & 0.194 \\
        \hline
    \end{tabular}
\end{table}
To assess the effect of resource density, we repeated the \(503\times503\) experiments at
\(\rho=0.005\), \({0.010}\), \(0.15\), \(0.30\), and \(0.45\), keeping the evolutionary protocol and resource-environment classes fixed. Figure~\ref{fig:density_results_503} summarizes the resulting best evolved trajectories,
final-population step-length distributions, and the model-selection score \(\Gamma\). Across all densities and environments, \(\Gamma\) remained positive, indicating that the intermittent-search model more accurately describes the evolved trajectories than the Lévy-walk model. The trajectory panels show that increasing density changes the spatial organization of movement, while the step-length panels show corresponding density-dependent shifts in the evolved genome. In particular, higher densities tend to concentrate step-length distributions more strongly at short values, although the strength of this effect depends on the underlying resource environment.

\paragraph{{Fixed-genome cue-off evaluation and coverage-term ablation}}
{To assess how the sensory cue affects the expressed trajectories, we performed an evaluation-only ablation. Genomes evolved under the standard \(p_{\mathrm{cue}}=0.5\) condition were held fixed and re-evaluated with \(p_{\mathrm{cue}}=0\), without further evolution. Cue-on and cue-off evaluations used matched generations and identical resource fields. Removing the cue reduced \(\Gamma\) in the uniform, \(\mu=1.0\), \(\mu=1.5\), and \(\mu=2.0\) environments, but increased it for \(\mu=2.5\) and \(\mu=3.0\). Thus, the cue affects the model-comparison margin in an environment-dependent manner. Because the genomes were not re-evolved without the cue, this evaluation does not establish that IS-like dynamics evolve independently of sensing.}

To test whether the coverage term drives the model classification, we re-evolved the agents using \(F=\eta_E\), while retaining the remaining simulation and evolutionary settings.
\begin{table*}[t]
\centering
\caption{Coverage-term ablation at \(L=503\) and \(\rho=0.15\). Values are mean \(\pm\) standard deviation over the final 20 best-agent generations. They represent correlated late-generation best-agent evaluations from one evolutionary run per objective, rather than independent evolutionary replicates.}
\label{tab:fitness_ablation}
\small
\setlength{\tabcolsep}{5pt}
\begin{tabular}{lcccc}
\toprule
{Environment}
& {\(\eta_C\), baseline \(F=\eta_E\eta_C\)}
& {\(\eta_C\), ablation \(F=\eta_E\)}
& {Retained \(\eta_C\) (\%)}
& {\(\Gamma\), ablation \(F=\eta_E\)} \\
\midrule
{Uniform}   & {\(0.9278\pm0.0065\)} & {\(0.9283\pm0.0048\)} & {\(100.1\%\)} & {\(0.0190\pm0.0013\)} \\
{\(\mu=1.0\)} & {\(0.9282\pm0.0057\)} & {\(0.9232\pm0.0055\)} & {\(99.5\%\)}  & {\(0.0180\pm0.0021\)} \\
{\(\mu=1.5\)} & {\(0.8997\pm0.0053\)} & {\(0.8920\pm0.0103\)} & {\(99.1\%\)}  & {\(0.0216\pm0.0034\)} \\
{\(\mu=2.0\)} & {\(0.8585\pm0.0083\)} & {\(0.8489\pm0.0073\)} & {\(98.9\%\)}  & {\(0.0313\pm0.0049\)} \\
{\(\mu=2.5\)} & {\(0.8129\pm0.0076\)} & {\(0.8105\pm0.0112\)} & {\(99.7\%\)}  & {\(0.0373\pm0.0100\)} \\
{\(\mu=3.0\)} & {\(0.7834\pm0.0118\)} & {\(0.7422\pm0.0154\)} & {\(94.7\%\)}  & {\(0.0509\pm0.0097\)} \\
\bottomrule
\end{tabular}
\end{table*}
Removing \(\eta_C\) from the objective did not eliminate high coverage efficiency or the positive IS--LW fit difference. The coverage term may still affect the degree of redundancy, particularly for \(\mu=3.0\), but it is not solely responsible for the classification result.

\begin{table}[t]
\centering
\caption{Fixed-genome cue-off evaluation at \(L=503\) and \(\rho=0.15\). Genomes evolved under the standard \(p_{\mathrm{cue}}=0.5\) condition were re-evaluated with \(p_{\mathrm{cue}}=0\), without further evolution. The cue-on and cue-off \(\Gamma\) values are the mean \(\pm\) standard deviation over the final \(500\) matched best-agent generations. The final column reports the mean adjusted LW coefficient of determination in the cue-off condition. These results are separate from the five-run aggregates in Table~\ref{tab:model_comparison_summary}.}
\label{tab:cue_off_evaluation}
\small
\setlength{\tabcolsep}{4pt}
\begin{tabular}{lccc}
\toprule
{Environment}
& {\(\Gamma_{\mathrm{cue\,on}}\)}
& {\(\Gamma_{\mathrm{cue\,off}}\)}
& {\(\bar{R}^{2}_{\mathrm{LW,cue\,off}}\)} \\
\midrule
{Uniform}     & {\(0.0175\pm0.0014\)} & {\(0.0010\pm0.0001\)} & {\(0.9988\)} \\
{\(\mu=1.0\)} & {\(0.0185\pm0.0017\)} & {\(0.0029\pm0.0008\)} & {\(0.9969\)} \\
{\(\mu=1.5\)} & {\(0.0204\pm0.0024\)} & {\(0.0059\pm0.0024\)} & {\(0.9933\)} \\
{\(\mu=2.0\)} & {\(0.0326\pm0.0046\)} & {\(0.0145\pm0.0062\)} & {\(0.9818\)} \\
{\(\mu=2.5\)} & {\(0.0251\pm0.0060\)} & {\(0.0267\pm0.0175\)} & {\(0.9720\)} \\
{\(\mu=3.0\)} & {\(0.0424\pm0.0085\)} & {\(0.0797\pm0.0483\)} & {\(0.9181\)} \\
\bottomrule
\end{tabular}
\end{table}

\subsection*{Synthetic validation of the classifier}
\label{app:synthetic_validation}
\begin{figure}[t] 
\centering
\includegraphics[width=0.49\textwidth]{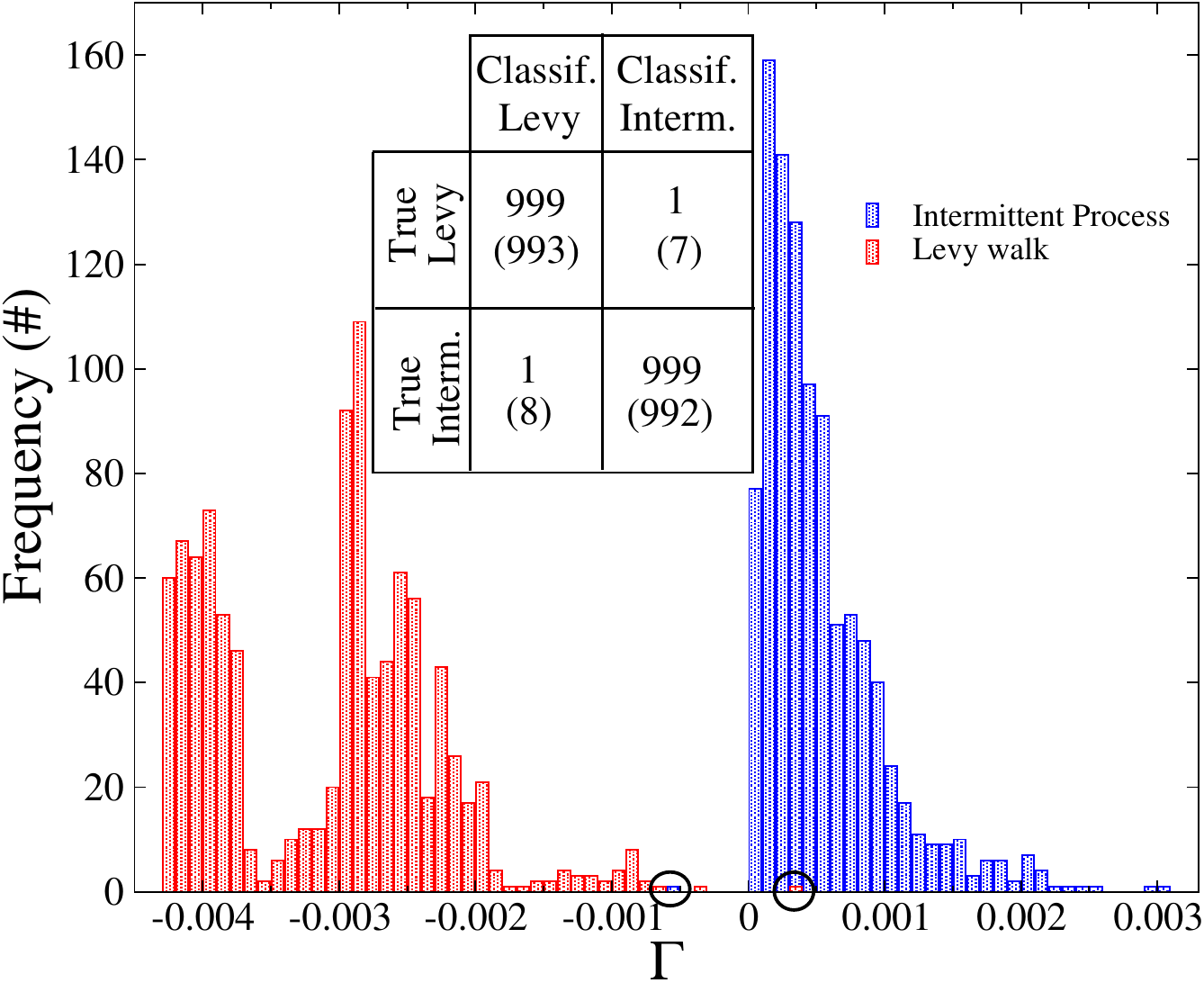}\hfill
\includegraphics[width=0.49\textwidth]{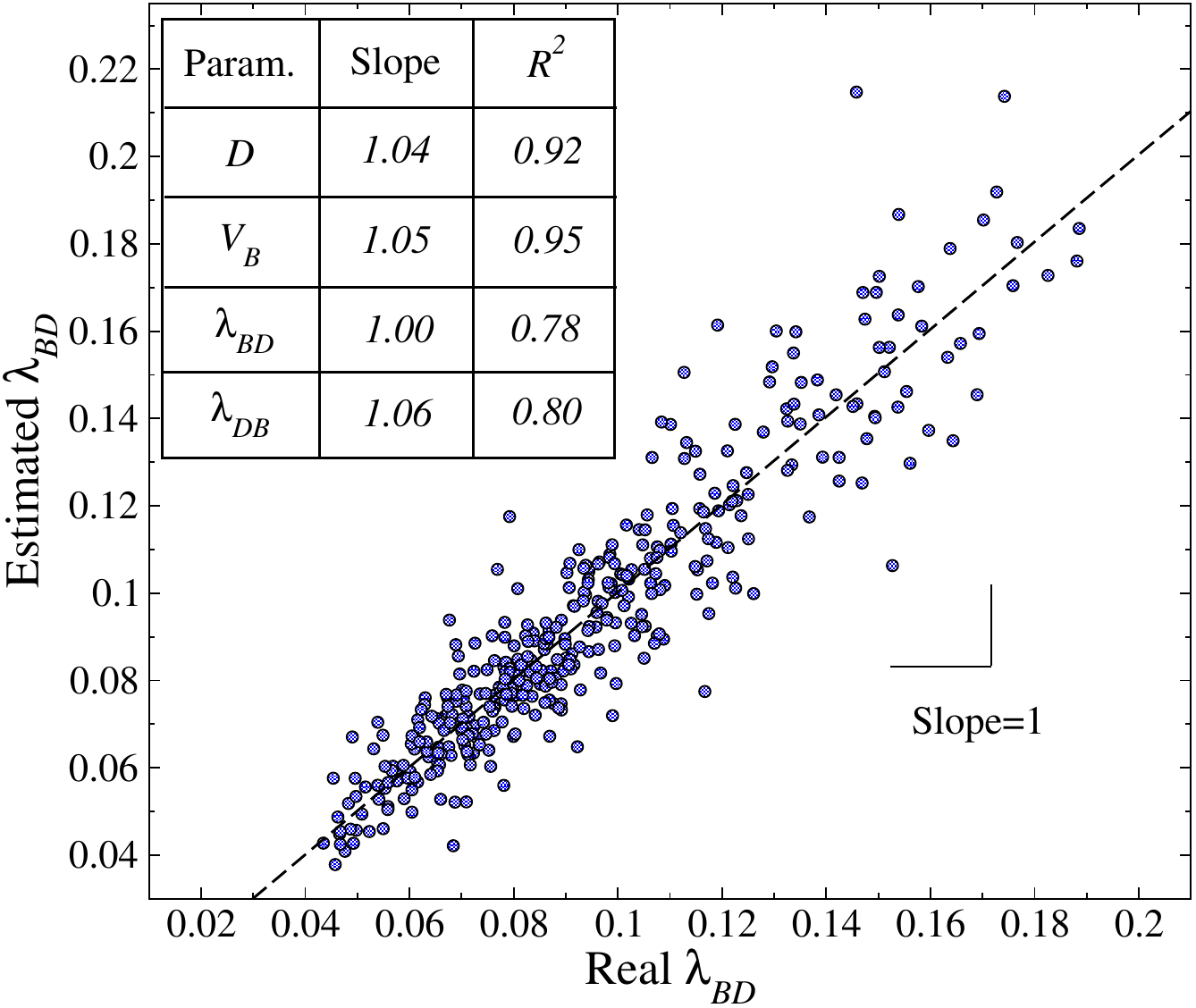}
\caption{\protect Synthetic validation of the moment-based classifier and intermittent-search parameter estimation. Left: distributions of the model-comparison score \(\Gamma=\bar{R}^{2}{\mathrm{IS}}-\bar{R}^{2}{\mathrm{LW}}\) for 1000 synthetic L'evy-walk trajectories and 1000 synthetic intermittent-search trajectories. Positive values favor IS, whereas negative values favor LW. The confusion matrix shows that 999 of 1000 trajectories from each generating model were classified correctly. Right: estimated versus generating values of \(\lambda_{BD}\) for synthetic IS trajectories. The dashed line marks exact recovery. The inset summarizes the regression slopes and coefficients of determination for \(D\), \(v_B\), \(\lambda_{BD}\), and \(\lambda_{DB}\).}
\label{fig:synthetic_validation}
\end{figure}
Figure~\ref{fig:synthetic_validation} reports classification results for synthetic LW and IS trajectories with known generating models, together with parameter-recovery results for the IS model. The parameter-recovery coefficients of determination range from \(0.78\) to \(0.95\), with fitted slopes between \(1.00\) and \(1.06\).
\subsection*{Temporal structure of evolved search trajectories}
\label{app:waiting_time_statistics}
\begin{table}[t]
\centering
\caption{\protect
Waiting-time statistics for the best evolved agents.
Fitness and efficiency values - cf. Eqs.~(\ref{eq:fitness}) and (\ref{eq:fitness_expanded}) - are reported as mean \(\pm\) standard deviation across grid sizes . Upper-tail spread is summarized by the median across grid sizes of \(q_{0.95}/\mathrm{median}\). Return intervals quantify recurrence to previously visited sites, resource-hit intervals quantify the temporal spacing between resource encounters, and new-site discovery intervals quantify the timing of first visits to previously unvisited sites.}
\label{tab:waiting_time_summary}
\renewcommand{\arraystretch}{1.15}
\setlength{\tabcolsep}{4.5pt}
\begin{tabular}{lcccccc}
\toprule
Env.
& \( {\mathcal F}\)
& \(\eta_E\)
& \(\eta_C\)
& Return
& Hit
& New-site \\
& 
& 
& 
& \(q_{0.95}/\mathrm{mdn.}\)
& \(q_{0.95}/\mathrm{mdn.}\)
& \(q_{0.95}/\mathrm{mdn.}\) \\
\midrule
\(\mu=1.0\) & 0.40 $\pm$ 0.02 & 0.43 $\pm$ 0.01 & 0.92 $\pm$ 0.02 & 25.4 & 3.0 & 2.0 \\
\(\mu=1.5\) & 0.39 $\pm$ 0.01 & 0.43 $\pm$ 0.01 & 0.90 $\pm$ 0.01 & 18.6 & 3.5 & 2.0 \\
\(\mu=2.0\) & 0.40 $\pm$ 0.01 & 0.46 $\pm$ 0.01 & 0.86 $\pm$ 0.01 & 81.5 & 6.0 & 2.0 \\
\(\mu=2.5\) & 0.37 $\pm$ 0.02 & 0.48 $\pm$ 0.02 & 0.81 $\pm$ 0.01 & 55.6 & 6.0 & 2.1 \\
\(\mu=3.0\) & 0.37 $\pm$ 0.05 & 0.49 $\pm$ 0.04 & 0.76 $\pm$ 0.06 & 42.9 & 5.0 & 3.0 \\
Uniform & 0.40 $\pm$ 0.02 & 0.43 $\pm$ 0.02 & 0.93 $\pm$ 0.02 & 39.7 & 3.0 & 1.0 \\
\bottomrule
\end{tabular}
\end{table}
\begin{figure}[t]
    \centering
    \includegraphics[width=\textwidth]{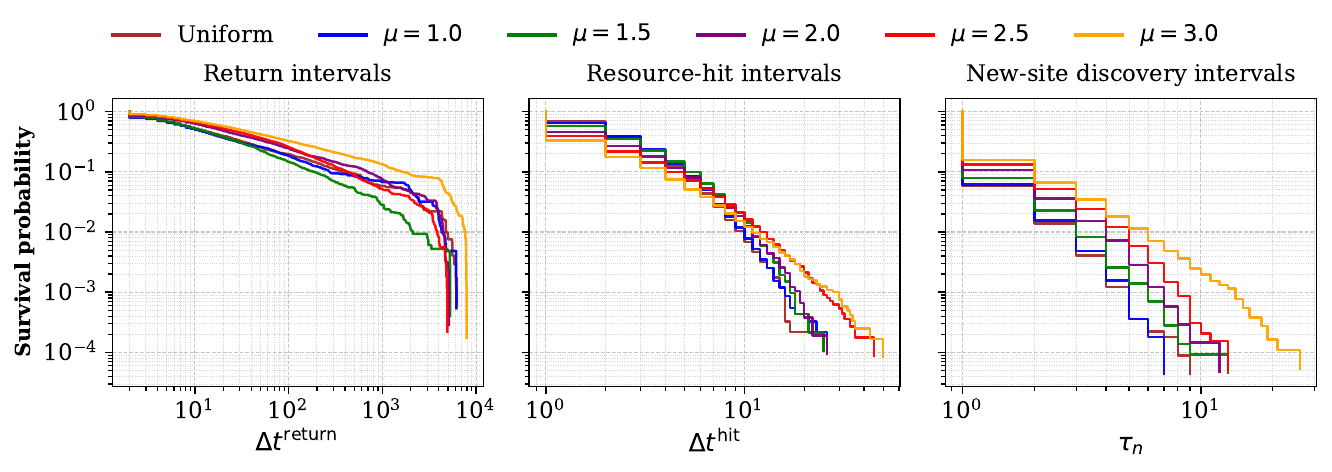}
    \caption{\protect
    Empirical survival distributions of waiting-time observables in evolved search trajectories. \textbf{Left}: return intervals \(\Delta t^{\mathrm{return}}\), measuring the time between successive visits to the same previously visited site.
    \textbf{Middle}: resource-hit intervals \(\Delta t^{\mathrm{hit}}\), measuring the time between successive resource encounters. \textbf{Right}: new-site discovery intervals \(\tau_n\), measuring the waiting time between first visits to successive previously unvisited sites.} 
    \label{fig:waiting_time_survival}
\end{figure}

To quantify temporal organization beyond trajectory visualization, we computed three waiting-time observables from the best evolved trajectories. Let \(\mathbf r_t\in\mathbb{Z}^2\), \(t=0,\;\ldots,\; T_{\mathrm{life}}\), denote the lattice position of an agent during one lifetime. For a visited site \(\mathbf z\), define the ordered set of visit times $ \mathcal V(\mathbf z)=\{t:\mathbf r_t=\mathbf z\}=\{t^{(\mathbf z)}_1<\cdots\;<\:t^{(\mathbf z)}_{m_{\mathbf z}}\}$.
Return intervals are the successive gaps between visits to the same site, $\Delta t^{\mathrm{return}}_{\mathbf z,k}=t^{(\mathbf z)}_{k+1}-t^{(\mathbf z)}_k,\;k=1,\;\ldots,\;m_{\mathbf z}-1$, pooled over all sites with \(m_{\mathbf z}\ge 2\). These intervals quantify recurrence to already visited locations.

If \(h_1<\cdots<h_{N_{\mathrm{hit}}}\) are the resource encounter times, resource-hit intervals are defined as $\Delta t^{\mathrm{hit}}_j=h_{j+1}-h_j, \; j=1,\;\ldots,\; N_{\mathrm{hit}}-1$. These intervals quantify the temporal spacing of successful foraging events. Finally, let \(t_n^{\mathrm{new}}\) be the first time at which the \(n\)-th distinct lattice site has been discovered, $t_n^{\mathrm{new}}=\min\left\{t:\left|\{\mathbf r_0,\ldots,\mathbf r_t\}\right|=n \right\}$. The new site discovery interval is then $\tau_n=t_{n+1}^{\mathrm{new}}-t_n^{\mathrm{new}}$. This definition follows the visitation-statistics convention in which \(\tau_n\) measures the waiting time to discover a new site after \(n\) distinct sites have already been visited. It is therefore distinct from return intervals, which measure recurrence to previously visited sites. For each waiting-time observable \(Y\), with observed values, \(Y_1,\ldots,Y_{n_Y}\), we computed the empirical survival function
\begin{equation}
\widehat S_Y(y)=\frac{1}{n_Y}\sum_{j=1}^{n_Y}\mathbf 1(Y_j\ge y)\;.
\end{equation} 
Thus, \(\widehat S_Y(y)\) is the fraction of observed intervals whose duration is at least \(y\). Slowly decaying survival curves indicate broad upper tails, corresponding to many short intervals together with rarer long gaps. We summarized this upper-tail spread by
 $\frac{q_{0.95}(Y)}{\mathrm{median}(Y)}$, where \(q_{0.95}(Y)\) is the empirical 95th percentile.

Table~\ref{tab:waiting_time_summary} summarizes these quantities after aggregation across the three grid sizes. Return intervals show the strongest right-skewed structure, resource-hit intervals are moderately right-skewed, and new-site discovery intervals remain comparatively narrow. This distinction is important: the strongest intermittency is observed in recurrence and resource-encounter timing, whereas spatial discovery proceeds at a comparatively regular rate.

Figure~\ref{fig:waiting_time_survival} clarifies how temporal intermittency differs across observables. Since the survival function \(\widehat S_Y(y)\) gives the fraction of intervals with duration at least \(y\), more slowly decaying curves indicate broader upper tails and a greater incidence of rare long intervals. Return intervals show the broadest tails, resource-hit intervals show intermediate tails, and new-site discovery intervals are comparatively narrow. This indicates that the strongest intermittency in the evolved trajectories appears in recurrence to previously visited sites and in the timing of resource encounters, whereas discovery of new sites proceeds on a more regular timescale.

\section{Analytical expressions for the second and fourth moments of an intermittent process}
\label{append:moments}
To select between L\'evy and Intermittent strategies, we use the derivations published recently in \cite{2bzm-t9k1} and implemented as an open source Python library (cf.~\cite{BHANDARI2025102334}).
These results enable us to distinguish between a planar uniform LW process, as described in Section \ref{classification} (cf.~Eq.~(\ref{eq:lw_pdf_revised})), and an IS model.

As mentioned in the main text, for the IS model, we can derive the analytical expressions of the second and the fourth moments.
Namely, the second moment is given by \cite{2bzm-t9k1}:
\begin{equation}
    \langle \mathbf{r}^2(t)\rangle = C_1^{(2)} t-C_{2}^{(2)}(1-e^{-\beta\alpha t}) \,
\label{eq:secondmomentinterm}
\end{equation}
with the values of the constants given by
\begin{eqnarray}
    C_1^{(2)} &=& 2 \frac{1-\alpha}{\alpha} D_I + 4\alpha D, \cr 
    C_2^{(2)} &=& 2 \frac{1-\alpha}{\alpha^2 \beta}D_I,\nonumber
\end{eqnarray}
where $\beta = \lambda_{BD}+\lambda_{DB}$, $\alpha =\lambda_{BD}/\beta$, and  $D_I=v_B^2/\beta$,
with \(\lambda_{BD}\) and \(\lambda_{DB}\) the transition rates from the ballistic to the diffusive regime and vice-versa, respectively. 

The fourth moment is given by 
\begin{eqnarray}
 \langle \mathbf{r}^4(t)\rangle
    &=& C_1^{(4)}t^2+C_2^{(4)} t-C_3^{(4)} - C_4^{(4)}e^{-\beta t}  +  \left ( C_5^{(4)}t^2 + C_6^{(4)}t -C_7^{(4)} \right ) e^{-\beta\alpha t} \label{eq:fourthmomentinterm}
\end{eqnarray}
with
\begin{eqnarray}
    C_1^{(4)}  &=&  8\alpha^2 \left( 2D+ 
    \frac{1-\alpha}{\alpha^2} D_I    \right)^2 ,\nonumber\\ 
    & & \cr
    C_2^{(4)}  &=& 8\frac{1-\alpha}{\alpha^3 \beta^2}
                 \left[ \left( D-\frac{\alpha+1}{2\alpha^2} D_I \right)^2 - \frac{5}{8\alpha^4} D_I^2 \right], \nonumber\\
    & & \cr
    C_3^{(4)}  &=& 4\frac{1-\alpha}{\alpha^4 \beta^4} \left[ \left ( 2D-\frac{\alpha^2+\alpha+1}{\alpha^3} D_I \right )^2 - \frac{6\alpha^3-6\alpha^2-\alpha-2}{2\alpha^6} \right]  , \nonumber\\
    & & \cr
    C_4^{(4)}  &=& 32(1-\alpha) \alpha\beta\left ( 2D+\frac{1}{1-\alpha} D_I    \right )^2  ,
    \cr
    & & \cr
    C_5^{(4)} &=& 4\frac{1-\alpha}{\alpha^2} D_I^2,\nonumber\\
    & & \cr
    C_6^{(4)}  &=&  \frac{16}{\alpha^2 \beta} D_I^2, \cr
    & & \cr
    C_7^{(4)}  &=& \frac{8}{\alpha^2\beta^2} D_I \left ( 8D+\frac{11\alpha^2-12\alpha+3}{\alpha^2 (1-\alpha)} D_I \right )^2.
    \nonumber
\end{eqnarray} 
These are the expressions of the moments used to compute \(m_k^{\mathrm{mod}}\) in Eq.~\eqref{eq:loss_revised} for the IS model.

The derivation of these analytical expressions is given in detail in Ref.~\cite{2bzm-t9k1} and can be summarized as follows:
\begin{itemize}
    \item Main idea: The IS process is written as a two-state planar stochastic process, where a walker alternates between the ballistic phase (phase "$+$" below), with speed \(v_B\), and a diffusive phase \(D\) (phase "$-$" below), with diffusivity \(D\). 
    \item Main aim: derive an analytical expression for the probability density function for each position $\mathbf r$ and time $t$, expressed as the sum of densities defined for each phase, namely
    \[ P(\mathbf r,t)=P_+(\mathbf r,t)+P_-(\mathbf r,t)\].
    \item One then introduce auxiliary quantities, namely densities \(\nu_{\pm}(\mathbf r,t)\) describing the switching between phases, residence-time densities in each phase, given by \(\psi_{\pm}(t)\), and conditional displacement densities during each phase, denoted by \(w_{\pm}(\mathbf r,t)\). 
    \item The switching densities \(\nu_{\pm}(\mathbf r,t)\) describe the flux of trajectories that arrive at \((\mathbf r,t)\) while switching from the opposite phase into phase \(\pm\). With initial phase probabilities \(c_\pm\), the renewal equations take the form $\nu_{\pm}(\mathbf r,t)=c_{\mp}q_{\mp}(\mathbf r,t)+\int_0^t\!\mathrm dt'\int_{\mathbb R^2}\!\mathrm d\mathbf r'\,q_{\mp}(\mathbf r',t')\nu_{\mp}(\mathbf r-\mathbf r',t-t')$, and $P_{\pm}(\mathbf r,t)=c_{\pm}Q_{\pm}(\mathbf r,t)+
    \int_0^t\!\mathrm dt'\int_{\mathbb R^2}\!\mathrm d\mathbf r'\,Q_{\pm}(\mathbf r',t')\nu_{\pm}(\mathbf r-\mathbf r',t-t')$,
    where $q_{\pm}(\mathbf r,t)=w_{\pm}(\mathbf r,t)\psi_{\pm}(t)$, and $Q_{\pm}(\mathbf r,t)=w_{\pm}(\mathbf r,t)\int_t^\infty \psi_{\pm}(t')\,\mathrm dt' $.

    \item The renewal equations are then transformed into the Fourier--Laplace space, where the convolutions become products. 
    After algebraic manipulation, the transformed
    probability density function for each position and time reads
    \[P(\mathbf k,s)=\frac{Q_{+}(\mathbf k,s)\left[c_{+}+c_{-}q_{-}(\mathbf k,s)\right]+
    Q_{-}(\mathbf k,s)\left[c_{-}+c_{+}q_{+}(\mathbf k,s)\right]}{1-q_{+}(\mathbf k,s)q_{-}(\mathbf k,s)}.
    \]
    \item The model is then specified by exponential residence-time densities, $\psi_+(t)=\lambda_{BD} e^{-\lambda_{BD} t},\; \psi_-(t)=\lambda_{DB} e^{-\lambda_{DB} t}$,
    where \(\lambda_{BD}\) and \(\lambda_{DB}\) are the transition rates from phase $+$ to phase $-$ and from $-$ to $+$ respectively. Substituting these expressions in the functions $q_{\pm}(\mathbf k,s)$ and $Q_{\pm}(\mathbf k,s)$, one obtains after algebraic manipulation
    \[P(\mathbf k,s)=\frac{c_B(s+\lambda_{DB}+Dk^2)
    +c_D\sqrt{(s+\lambda_{BD})^2+v_B^2k^2}
    +c_B\lambda_{BD}+c_D\lambda_{DB}}{(s+\lambda_{DB}+Dk^2)
    \sqrt{(s+\lambda_{BD})^2+v_B^2k^2}-\lambda_{BD}\lambda_{DB}},\] with \(c_B=\frac{\lambda_{DB}}{\lambda_{BD}+\lambda_{DB}}\) and \(c_D=\frac{\lambda_{BD}}{\lambda_{BD}+\lambda_{DB}}\).
    
    \item Finally, the second and fourth moments are obtained by differentiating this propagator at \(\mathbf k=\mathbf 0\) and then applying the inverse Laplace transform $\mathcal L^{-1}$:
    \[\langle \mathbf r^2(t)\rangle=\mathcal L^{-1}\!\left[
    -\left(\frac{\partial^2}{\partial k_x^2}+
    \frac{\partial^2}{\partial k_y^2}\right)P(\mathbf k,s)\bigg|_{\mathbf k=\mathbf 0}\right],\]
    and \[\langle \mathbf r^4(t)\rangle=\mathcal L^{-1}\!\left[\left(\frac{\partial^4}{\partial k_x^4}
    +2\frac{\partial^4}{\partial k_x^2\partial k_y^2}
    +\frac{\partial^4}{\partial k_y^4}\right)P(\mathbf k,s)\bigg|_{\mathbf k=\mathbf 0}\right].\]
\end{itemize}
For the LW model, since the velocity $v_L$ enters the definitions of the moments only trivially, it is used here as a scaling parameter. The two remaining parameters, $\tau_0$ and $\gamma$, are estimated with a grid search algorithm. Further details on the derivations and their implementation in the {\it IntLevPy} library are given in Refs.~\cite{2bzm-t9k1, BHANDARI2025102334}.

\end{document}